\documentclass[12pt]{iopart}

\expandafter\let\csname equation*\endcsname\relax

\expandafter\let\csname endequation*\endcsname\relax

\usepackage{amsmath}
\usepackage{amssymb}
\usepackage{graphicx}

\usepackage[caption=false]{subfig}

\usepackage{epstopdf}
\usepackage{verbatim}
\usepackage{hyperref}
\hypersetup{colorlinks,linkcolor=blue,urlcolor=blue,citecolor=magenta}

\newcommand{\HH}{\mathcal{H}}

\def\be{\begin{equation}}
\def\ee{\end{equation}}
\def\erf{\eqref}

\newcommand{\dd}{\mathrm{d}}

\newcommand{\vev}[1]{\langle #1 \rangle}
\newcommand{\fR}   {f^{\text{R}}}
\newcommand{\fL}   {f^{\text{L}}}
\newcommand{\ud}          {\mathrm d}
\newcommand\eps           {\varepsilon}
\newcommand{\Li}[1]{\mathrm{Li}_2\left( #1 \right)}
\newcommand{\twomat}[4]{\left(\begin{array}{cc} #1 & #2 \\ #3 & #4\end{array}\right)}
\newcommand{\unity}{\ensuremath{{\rm 1 \hspace{-0.27778em} l}{}}}

\newcommand{\sig}{\sigma}
\newcommand{\TR}   {T_{\text{R}}}
\newcommand{\TL}   {T_{\text{L}}}

\newcommand{\phiR}   {\phi^\text{R}}
\newcommand{\phiL}   {\phi^\text{L}}
\newcommand{\psiR}   {\psi^\text{R}}
\newcommand{\psiL}   {\psi^\text{L}}
\newcommand\fii           {\varphi}

\usepackage{indentfirst}
\usepackage{color}

\begin{document}

\frenchspacing

\title[Temperature driven quenches in the Ising model]{Temperature driven quenches in the Ising model: appearance of negative R\'enyi mutual information}

\author{M\'arton Kormos}

\address{
BME-MTA Statistical Field Theory Research Group, Institute of Physics,
Budapest University of Technology and Economics, H-1111 Budapest, Hungary}
\ead{kormos@eik.bme.hu}

\author{Zolt\'an Zimbor\'as}
\address{Dahlem Center for Complex Quantum Systems, Freie Universit\"at Berlin, 14195 Berlin, Germany}
\address{Wigner Research Centre for Physics, Hungarian Academy of Sciences, P.O. Box 49, H-1525 Budapest, Hungary}
\ead{zimboras@gmail.com}

\vspace{10pt}

\begin{abstract}
We study the dynamics of the transverse field Ising chain after a local quench in which two independently thermalised chains are joined together and are left to evolve unitarily. In the emerging non-equilibrium steady state the R\'enyi mutual information with different indices are calculated between two adjacent segments of the chain, and are found to scale logarithmically in the subsystem size. Surprisingly, for R\'enyi indices $\alpha>2$ we find cases where the prefactor of the logarithmic dependence is negative. The fact that the naively defined R\'enyi mutual information might be negative has been pointed out before, however, we provide the first example for this scenario in a realistic many-body setup. Our numerical and analytical results indicate that in this setup it can be negative for any index $\alpha>2$ while it is always positive for $\alpha <2.$ Interestingly, even for $\alpha>2$ the calculated prefactors show some universal features: for example, the same prefactor is also shown to govern the logarithmic time dependence of the R\'enyi mutual information before the system relaxes locally to the steady state. In particular, it can decrease in the non-equilibrium evolution after the quench.
\end{abstract}

%\maketitle

\section{Introduction}

In the past decade, the study of correlations between subsystems in lattice models and in field theories has allowed for a deeper understanding of the physics of many-body systems, in particular in relation to quantum criticality~\cite{CC09,Amico2008,Calabrese2009a,Eisert2010,Laflorencie2015}, equilibration~\cite{GogolinEisert16,Alba2016}, and topological order \cite{Wen13, SavaryBalents16}. In pure states (e.g. in ground states), the correlation between two complementary subsystems is entirely quantum mechanical and can be measured by the entanglement entropy.  The ground state entanglement entropy for one dimensional gapped local Hamiltonians was proved to obey an area law \cite{H07, Brandao13}, while for critical models that can be described by a conformal field theory (CFT) it was shown to grow logarithmically in the subsystem size \cite{HLW94, Vidal03, CC04, Korepin04}. More exotic scaling behaviour was also found in other types of gapless models \cite{FZ05,Irani10,MS16,SUZK16}.

The use of entanglement entropy as a correlation measure is restricted to pure states and a bipartite setting. When the system is in a mixed state on a bipartite Hilbert space $\HH_A \otimes \HH_B$, the correlation between subsystems $A$ and $B$ can be characterised by the mutual information (MI),  
\begin{equation} \label{eq:mut_inf}
I(A:B)=S(\rho_A) + S(\rho_B) - S(\rho_{AB})\,,
\end{equation}
where  $\rho_{A}$ and $\rho_B$ are the reduced density matrices of the subsystems $A$ and $B$, and $S$ denotes the von Neumann (or entanglement) entropy,
\begin{equation}
S(\rho)= - \Tr \rho \log \rho\,.
\end{equation}
The MI has many nice properties. It is positive due to the subadditivity of the von Neumann entropy, and is zero if and only if $\rho_{AB}=\rho_A \otimes \rho_B$, i.e. when the state is uncorrelated. An operational interpretation of the MI is that it measures (asymptotically) the minimal amount of noise needed to erase the correlation in the state by turning it into a product state \cite{GPW05}. This is related to the fact that the MI is equal to the relative entropy (or quantum Kullback--Leibler divergence) between $\rho_{AB}$ and $\rho_A \otimes \rho_B$. 
The relative entropy, defined as
\begin{equation}
D(\rho \, || \, \sigma)= \Tr \rho(\log\rho - \log \sigma)\,,
\end{equation} 
is a measure of distinguishability between two quantum states $\rho$ and $\sigma$. There has been an increasing activity on the relative entropy in field theory, see \cite{Ruggiero2016} and references therein. The mentioned relation to the MI can be shown by the following standard derivation
\begin{align*}
D (\rho_{AB} \, || \, \rho_A \otimes \rho_B)&= \Tr_{\HH_A \otimes \HH_B} \rho_{AB}(\log \rho_{AB}-  \log \rho_A \otimes \rho_B)\\
&= {-}S(\rho_{AB})  {-} \Tr_{\HH_A \otimes \HH_B} ( \rho_{AB} \log \rho_A \otimes \unity_B  +  \rho_{AB} \log \unity_A \otimes \rho_B)\\
&=-S(\rho_{AB}) -  \Tr_{\HH_A} \rho_A \log \rho_A    - \Tr_{ \HH_B} \rho_B \log \rho_B=I(A:B)\,.
\end{align*} 
Alternatively, the mutual information can also be charaterised as the following minimum
\begin{equation} \label{eq:min_MI}
I(A:B)= \min_{\sigma_B} D (\rho_{AB} \, || \, \rho_A \otimes \sigma_B)\, ,
\end{equation}
where $\sigma_B$ is any density matrix on the Hilbert space $\HH_B$ of subsystem $B$.

It was shown that in finite temperature Gibbs states of local Hamiltonians a strict area law holds for the mutual information \cite{WVHC08, BKE15}.  Until now, the only examples of a violation of the area law in the MI outside the zero temperature regime was found to appear in non-equilibrium steady states (NESS) of
spin chains \cite{EZ14}.
 %the XX model \cite{EZ14}, followed by a subsequent study for the case of the XY chain \cite{ABZ14}. 
 These states can be written as Gibbs states of infinite-range Hamiltonians, and thus the theorems of Refs. \cite{WVHC08, BKE15} which use the local structure of the interactions do not apply.

Besides the von Neumann entropy and the quantities directly derived from it (such as the mutual information and the topological entanglement entropy \cite{Hamma05, KP06, LevinWen06}), the R\'enyi entropies were also shown to play an important role in many-body physics. The R\'enyi entropy with index $\alpha$  is defined as 
\begin{equation}
S^{(\alpha)}(\rho)=\frac{1}{1-\alpha} \log \Tr(\rho^{\alpha})\,.
\end{equation}
Note that in the $\alpha \to 1$ limit we recover the von Neumann entropy.  R\'enyi entropies with integer $\alpha$ indices (and with $\alpha > 1$) appear as natural quantities in conformal field theory through the replica approach \cite{CC04,CC06, Furukawa2009,Calabrese2009}. These quantities are also easier to calculate than the von Neumann entropy in various analytical and numerical settings \cite{HGKM10,Grover13,ALT14,Palmai16}, and one can use them to detect criticality \cite{CC09} and topological order \cite{FHHW09}.   Moreover, the experimental determination of R\'enyi entropies also seems more feasible \cite{Cardy2011,AD12, islam2015,Kaufman2016,Alba2016a}. Following this line of studies, as a natural generalisation of Eq.~\eqref{eq:mut_inf}, also the R\'enyi mutual information with index $\alpha$ was introduced as
\begin{equation} \label{eq:Reny_Mut_Inf}
I^{(\alpha)}(A:B)=S^{(\alpha)}(\rho_A) + S^{(\alpha)}(\rho_B) - S^{(\alpha)}(\rho_{AB})\,.
\end{equation}
The R\'enyi mutual information has been shown to exhibit universal scaling behaviour in ground, excited and thermal states \cite{MKH10, Alba2010, SHKM11, IIKM13, Coser2014b}. Furthermore, also for post-measurement states \cite{AR14}, in holographic settings \cite{Headrick10}, and  non-equilibrium scenarios \cite{AB14} certain universal features show up.

However, when discussing the extensive use of R\'enyi MI in many-body physics, it should be mentioned that, unlike the standard MI, this quantity has in general no operational meaning and may even be negative. Thus, in quantum information theory a different R\'enyi generalisation of MI is used. First, the relative entropy is generalised by defining the $\alpha$-R\'enyi divergences \cite{MDSFT13,Wilde2013},
\begin{equation}
D_1^{(\alpha)}(\rho \, || \, \sigma){=} \frac{1}{\alpha-1} \log \Tr \left( \rho^{\alpha}\sigma^{1-\alpha} \right) \,, \; \; \;D_2^{(\alpha)}(\rho \, || \, \sigma){=} \frac{1}{\alpha-1} \log \Tr \left( \sigma^{\frac{1-\alpha}{2\alpha}}\rho \sigma^{\frac{1-\alpha}{2\alpha}}\right)^{\alpha}\,,
\end{equation}
which are used, e.g., in state discrimination theory \cite{MO15}. Building on these divergences, by generalising Eq.~\eqref{eq:min_MI}, a regularised $\alpha$-R\'enyi mutual information can be introduced as
\begin{equation}\label{eq:Reg_Renyi}
I^{(\alpha)}_j(A:B)= \min_{\sigma_B} D^{(\alpha)}_j (\rho_{AB} \, || \, \rho_A \otimes \sigma_B)
\end{equation}
for $j=1,2$. These quantities are not only positive by definition, but have also other nice properties including operational interpretations \cite{Gupta2013,Cooney2014,BSW15,HT16}. 

Let us return to the R\'enyi MI defined by Eq.~\eqref{eq:Reny_Mut_Inf}. Despite the mentioned general problems with this quantity, for particular families of states it was found to be useful. For bosonic Gaussian states it was proved that the 2-R\'enyi mutual information is positive, or equivalently that the 2-R\'enyi entropy satisfies the subadditivity condition \cite{AGS12}. This allowed for the introduction of new types of correlation measures, e.g., steering quantifiers, which had no counterparts among quantities based on the conventional von Neumann entropy \cite{LHAW16}. Also in states appearing in the studied many-body scenarios the R\'enyi MI seemed to remain always positive and to show a behaviour very similar to that of the von Neumann MI, as discussed previously. Thus it has emerged as a natural quest to prove the positivity of this quantity together with possible area laws in a broad many-body context (see e.g. \cite{SDHS16}).

In the present paper we provide examples of naturally appearing many-body states for which the R\'enyi mutual information can be negative. In particular, we consider non-equilibrium steady states of the transverse field Ising model that emerge after joining two half-infinite chains thermalised at different temperatures, and calculate analytically and numerically the R\'enyi mutual information asymptotics finding cases where $I^{(\alpha)}$ is negative for $\alpha>2$. % \red{The R\'enyi entropies out of equilibrium were studied in \cite{Fagotti2008,Fagotti2010,Stephan2011}.}
%We also prove that for fermionic Gaussian states (which through the Jordan--Wigner transformation include the studied Ising NESS) the $2$-R\'enyi mutual information is always positive. 

The paper is organised in the following way. In Sec. \ref{sec:NESS} we introduce the physical setup: after briefly summarising the diagonalisation of the transverse field Ising spin chain, we show how the time evolution of correlations can be computed and present the building blocks of correlation functions in the NESS. In Sec. \ref{sec:RenyiMI} we discuss the main ideas behind the calculation of the R\'enyi MI and present exact closed form results for the prefactor of its logarithmic dependence on subsystem size in the NESS. We check our analytical expressions by comparing them to numerical calculations on finite lattices and we analyse the dependence of the prefactors  on the various parameters of the problem. In Sec. \ref{sec:MIevol} we turn to the numerical investigation of the non-equilibrium time evolution of the R\'enyi  MI after joining the two chains and provide evidence that after the initial transient and before relaxation to the steady state it depends logarithmically on time with the same prefactor that governs its spatial dependence in the NESS. We give our conclusions in Sec. \ref{sec:outlook}. The details of the analytical calculation of the R\'enyi MI is delegated to Appendix A.

\section{Transverse field Ising model: temperature driven quench and non-equilibrium steady state \label{sec:NESS}}

The system we study in this work consists of two half-infinite chains thermalised at different temperatures and brought to contact at time zero. This setup belongs to a more general scheme that is called in the literature ``cut and glue quench" or ``partitioning approach" which also includes the case of different chemical potentials (or magnetisation) on each side \cite{Antal1997}.
The non-equilibrium steady state was constructed for the two-temperature case in the XX and XY spin chains in \cite{Ho2000,Aschbacher2003,DeLuca2014a}. In \cite{Ogata2002,Platini2006} the spatial profile of the magnetisation density and current was determined in the XX spin chain, while the Ising spin chain was studied in \cite{Platini2005,DeLuca2013}. These systems can be mapped to free spinless fermions unlike the integrable XXZ spin chain investigated in \cite{Karrasch2013,DeLuca2014,Bertini2016b}. Continuum theories were investigated as well, including the free bosonic \cite{Doyon2014b} and fermionic \cite{Collura2014a,Collura2014} systems as well as integrable quantum field theories \cite{Doyon2012,Castro-Alvaredo2014,Castro-Alvaredo2016}. There is a growing body of results in conformal field theories where the energy density, the full distribution of the current and fluctuation relations in the NESS were obtained \cite{Bernard2012,Bhaseen2013,Bernard2014}, see \cite{Bernard2016a} for a review.

Along these lines, also the correlation between the left and right subsystems in the NESS was investigated. Considering quantum correlations, the logarithmic negativity between two adjacent subsystems was investigated within a CFT setting, and it was found to be the average of the two equilibrium negativity values \cite{EZ14b,Hoogeveen2014}.  The mutual information (measuring the total, classical and quantum, correlation) between two adjacent segments was shown to logarithmically violate a strict area law for the XX NESS \cite{EZ14}, and similar violation was numerically found for XY chains \cite{ABZ14}. In the present paper, we will follow this line of study by investigating the R\'enyi MI in the NESS of the transverse Ising chain. Before providing the main results, in
this section we shortly recall the basics of the Ising time evolution and the form of the emerging NESS.

\subsection{Diagonalising the Ising spin chain on a finite interval}\label{sec:2a}

The Hamiltonian of the transverse field Ising spin chain of length $N$ is 
\be
H=-\frac12 \sum_{j=1}^{N-1}\sigma_j^x\sigma_{j+1}^x-\frac12\sum_{j=1}^N h\sigma_j^z\,,
\ee
where $\sigma_j^\alpha$ are the Pauli matrices and we consider open boundary conditions. By the Jordan--Wigner transformation, $c_j=\prod_{k=1}^{j-1}(-\sig_k^z)\sig_j^-, c^\dag_j=\prod_{k=1}^{j-1}(-\sig_k^z)\sig_j^+,$ the Hamiltonian is mapped on that of free spinless fermions:
\be
\label{Hc}
H=-\frac12\sum_{j=1}^{N-1}\left[ c_j^\dag c_{j+1}+c_{j+1}^\dag c_j + c_j^\dag c^\dag_{j+1}+c_{j+1}c_j\right] -h\sum_{j=1}^N\left(c_j^\dag c_j-\frac12\right)\,,
\ee
where $\{c_j,c_k^\dag\}=\delta_{j,k},$ $\{c_j,c_k\}=\{c_j^\dag,c_k^\dag\}=0.$ It is useful to introduce the Majorana fermion operators
\be
a_{2j-1} = c_j+c^\dag_j\,,\qquad a_{2j} = i(c_j-c^\dag_j)\,.
\ee
The Hamiltonian is a bilinear form which can be diagonalised by a linear transformation leading to the fermionic mode operators
\begin{align}
\eta_k &= \frac12\sum_{j=1}^N \left[\phi_k(j)a_{2j-1}-i\psi_k(j)a_{2j}\right]\,,&
\eta_k^\dag &= \frac12\sum_{j=1}^N \left[\phi_k(j)a_{2j-1}+i\psi_k(j)a_{2j}\right]\,,\\
a_{2j-1}& = \sum_k \phi_k(j)(\eta^\dag_k+\eta_k)\,, &
a_{2j} &= -i\sum_k \psi_k(j)(\eta^\dag_k-\eta_k)
\end{align}
with the functions
\begin{subequations}
\begin{align}
\phi_k(j) &= A_k \sin(k j - \theta_k)\,,\\
\psi_k(j) &= -A_k \sin(k j)\,,
\end{align}
\end{subequations}
where $0<\theta_k<\pi$ is the Bogoliubov angle satisfying
\be
\label{eq:theta}
\tan\theta_k = \frac{\sin k}{h+\cos k}
\ee
and $A_k^{-2} = \sum_{j=1}^N\sin^2 (kj)$ is the normalisation.
%
%The inverse transformation is given by
%
%\be
%A_j = \sum_k \phi_k(j)(\eta^\dag_k+\eta_k)\,,\qquad B_j = \sum_k \psi_k(j)(\eta^\dag_k-\eta_k)\,.
%\ee
%
The modes satisfy $\{\eta_k,\eta^\dag_{k'}\}  = \delta_{k,k'},$ $\{\eta_k,\eta_{k'}\}=0,$ and in terms of them the Hamiltonian reads
\be
\label{Heta}
H=\sum_k \eps_k \eta_k^\dag\eta_k + \text{const.}
\ee
with the dispersion relation
\be
\label{eq:eps}
\eps_k= \sqrt{1+2h\cos k+h^2}\,.
\ee
%
%and the constant can be determined exploiting the fact that under a canonical transformation the trace of $H$ is invariant. 
In finite volume, the ``momentum" $k$ can only take quantised values according to the condition
\be
\label{quantcond}
k(N+1)-\theta_k = n\pi\,,\qquad n\in\mathbb{Z}\,.
\ee
%
%or alternatively,
%
%\be
%\frac{\sin[(N+1)k]}{\sin(Nk)} = -\frac1h\,.
%\ee
%

%%%%%%%%%%%%%%%%%%%%%%%%%%%%%%%%%%%%%%%%%%%%%

\subsection{Time evolution}

Our initial state corresponds to two independent, disjoint chains of length $N$ thermalised at different temperatures $\TL$ and $\TR,$ so the initial density matrix is
\be
\rho_0 = \rho_\text{L}(\TL)\otimes \rho_\text{R}(\TR)\,.
\ee
At time $t=0$ the two halves are joined and let evolve by the Hamiltonian $H$ of the chain of length $2N.$ In other words, we turn on the coupling between site 0 and site 1,
\be
H = H_\text{L} + H_\text{R} -\frac12\sigma^x_0\sigma^x_1 = \sum_k \eps_k\gamma^\dag_k\gamma_k+\text{const.}\,,
\ee
where $H_\text{L/R}$ are the Hamiltonian of the left and right chain of length $N,$ respectively, and the $\gamma_k$ are the mode operators that diagonalise $H.$

Let us denote the mode functions of $H_\text{R}$ by $\phi_q$ and $\psi_q,$ then the mode functions of $H_\text{L}$ are $\phiL_q(j)=\psiR_q(1-j)$ and $\psiL_q(j)=\phiR_q(1-j).$  As the first site of the full chain of length $2N$ has index $-N+1,$ the eigenfunctions of $H$  are $\varphi_k(j)=\phi_k(j+N)$ and $\chi_k(j)=\psi_k(j+N),$ where the momenta $\{k_n\}$ are quantised with $2N$ instead of $N$ in Eq. \erf{quantcond}. Note that the functional form of $\eps_k$ and $\theta_k$ are the same in all cases (we do not quench the Ising interaction or the transverse field). The Majorana operators on the full chain are related to the modes by
%related to the original spin operators, so we have
%
\be
a_{2j-1} = \sum_k \fii_k(j)(\gamma^\dag_k+\gamma_k)\,,\qquad 
a_{2j} = -i\sum_k \chi_k(j)(\gamma^\dag_k-\gamma_k)\,.
\ee

In order to compute the time evolution of correlation functions of spin operators, we first need to compute the building blocks given by the Majorana correlations $\vev{a_n(t)a_m(t)}.$
The time evolved operators in the Heisenberg picture are
\be
a_n(t)  =\sum_j\langle a_j|a_n(t)\rangle a_j\,,\\
\ee
where
\begin{subequations}
\label{ABcoeff}
\begin{align}
\vev{a_{2j-1}|a_{2n-1}(t)} &= \sum_k\varphi_k(j)\varphi_k(n)\cos(\eps_kt)\,,\\
\vev{a_{2j}|a_{2n}(t)} &= \sum_k\chi_k(j)\chi_k(n)\cos(\eps_kt)\,,\\
\vev{a_{2j-1}|a_{2n}(t)} = -\langle a_{2n} |a_{2j-1}(t)\rangle &= \sum_k\varphi_k(j)\chi_k(n)\sin(\eps_kt)\,.
\end{align}
\end{subequations}
In the infinite volume limit, $N\to\infty,$ the sum over $k$ turns into an integral. Dropping highly oscillating terms in the integrands we obtain
\begin{subequations}
\label{ABcoeff3}
\begin{align}
\vev{a_{2j-1}|a_{2n-1}(t)} = \vev{a_{2j}|a_{2n}(t)} & = \int_{-\pi}^\pi \frac{\dd k}{2\pi}\tilde\fii^*_k(j)\tilde\fii_k(n) \cos(\eps_kt)\,,\\
\vev{a_{2j-1}|a_{2n}(t)} = -\vev{a_{2n}|a_{2j-1}(t)}&= \int_{-\pi}^\pi \frac{\dd k}{2\pi} \tilde\fii^*_k(j)\tilde\chi_k(n) \sin(\eps_kt)\,,
\end{align}
\end{subequations}
where the infinite volume mode functions are
\be
\tilde\fii_k(j)=e^{-ikj+i\theta_k}\,,\qquad \tilde\chi_k(j)=-e^{-ikj}\,.
\ee
The time dependent Majorana two-point functions can be written as 
\be
\label{aacorr}
\langle a_n(t) a_m(t)\rangle = \sum_{j,l}
\langle a_j|a_n(t)\rangle \langle a_l | a_m(t') \rangle \,\langle a_ja_l\rangle_0\,.
\ee
Clearly, the initial correlations will be non-zero only if $j,l\ge1$ or $j,l\le0,$ so the correlation  function \erf{aacorr} splits into two parts corresponding to the contributions of the left and right half chains. 
%
%In the following we deal with the right chain (the left chain can be treated analogously). 
We now rewrite the Majorana operators in terms of the mode  operators $\eta_q$ diagonalising the left and right half chains and use for each half chain $\vev{\eta^\dag_q\eta^\dag_{q'}}_0= \vev{\eta_q\eta_{q'}}_0=0$ and $\vev{\eta^\dag_q\eta_{q'}}_0=\delta_{q,q'}f_q,$ where
\be
\label{eq:fq}
f_q = \frac1{1+e^{\eps_q/T_\text{L/R}}}
\ee
is the thermal Fermi--Dirac distribution function. Exploiting the completeness of the mode functions, we arrive at
\begin{subequations}
\begin{align}
\langle a_{2j-1}a_{2l-1}\rangle_0 &= \langle a_{2j}a_{2l}\rangle_0  = \delta_{j,l}\,,& j,l&\ge1\text{ or } j,l\le0\,,\label{aadelta}\\
\vev{ a_{2j-1}a_{2l}}_0 &= -\langle a_{2l}a_{2j-1}\rangle_0  = -i\sum_q\phi^\text{R}_q(j)\psi^\text{R}_q(l)(1-2f^\text{R}_q)\,,& j,l&\ge1\,,\\
\vev{ a_{2j-1}a_{2l}}_0 &= -\langle a_{2l}a_{2j-1}\rangle_0  = -i\sum_q\phi^\text{L}_q(j)\psi^\text{L}_q(l)(1-2f^\text{L}_q)\,,& j,l& \le 0\,.
%\vev{ A_jB_l}_0 &= -\langle B_lA_j\rangle_0  = \sum_q\tilde\phi_q(j)\tilde\psi_q(l)(1-2\fL_q)\,,\qquad j,l\le0\,.
\end{align}
\end{subequations}
%
%Here $f^\text{L/R}_q = (1+e^{\eps_q/T_\text{L/R}})^{-1}$ is the Fermi--Dirac distribution. 
In the $N\to\infty$ limit these become integral expressions.
%
%\begin{subequations}
%\label{AjBl2}
%\begin{align}
%\vev{ a_{2j-1}a_{2l}}_0 &= F^\text{(R)}(j-l) - F^\text{(R)}(j+l)\,, \quad j,l\ge1 \,,\\
%\vev{ a_{2j-1}a_{2l}}_0 &= F^\text{(L)}(j-l) - F^\text{(L)}[2-(j+l)]\,, \quad j,l\le0  \,,
%\end{align}
%\end{subequations}
%
%where
%
%\be
%F^{\text{(R/L)}}(j) = i\int_{-\pi}^\pi\frac{\dd q}{2\pi} (1-2f^\text{(R/L)}_q)\cos(qj-\theta_q)\,.
%\ee

Thanks to the orthonormality of the mode functions, the contribution of Eq. \erf{aadelta} yields a Kronecker $\delta_{n,m}$ in Eq. \erf{aacorr}, resulting in 
\be
\label{aasum}
\vev{a_n(t) a_m(t)} = \delta_{n,m}+
\sum_{j,l=-\infty}^\infty\Big[\vev{a_{2j-1}|a_n(t)}\vev{a_{2l} | a_m(t)}-\vev{a_{2j-1}|a_m(t)}\vev{a_{2l} | a_n(t)}\Big]\vev{a_{2j-1}a_{2l}}_0\,,
\ee
where both the coefficients and the initial correlations are written in the infinite $N$ limit in the form of integrals and any explicit dependence on $N$ disappeared. For numerical simulations, however, we use the finite $N$ expressions involving finite sums.

\subsection{Correlations and the GGE-like form of the asymptotic steady state}

The non-equilibrium steady state corresponds to the limit $t\to\infty$ in Eq. \erf{aasum} with $n,m$ fixed. In this limit, the gradients of all observables tend to zero resulting in a translationally invariant state. Expression \erf{aasum} can be greatly simplified  in this limit \cite{Marci}, which leads to the asymptotic correlations first derived in \cite{Aschbacher2003},
\begin{subequations}
\label{aaNESS}
\begin{align}
\label{ABas}
\vev{a_{2n-1}(t)a_{2m}(t)}_\text{NESS} &=  i\int_{-\pi}^\pi \frac{\dd k}{2\pi} \left(1-\fR_{k}-\fL_{k}\right) e^{i\theta_{k}}e^{ik(m-n)}\,,\\ \label{BBas}
\vev{a_{2n-1}(t)a_{2m-1}(t)}_\text{NESS}& = \vev{a_{2n}(t)a_{2m}(t)}_\text{NESS} =
\delta_{n,m}+\int_{-\pi}^\pi \frac{\dd k}{2\pi} (\fR_{k}-\fL_{k}) e^{ik(m-n)}\mathrm{sgn}(k)\,.
\end{align}
\end{subequations}
Thus, the NESS is a fermionic Gaussian state defined by the covariance matrix corresponding to the correlation functions \erf{aaNESS}. The Gaussianity of the NESS implies that it can be interpreted as a Gibbs state of an effective quadratic Hamiltonian,
\begin{equation}
\rho_{\text{NESS}}= \frac{1}{Z}  \exp (- \overline{\beta} \, H_{\text{eff}})\,,
\end{equation}
where $Z=\Tr[ \exp (- \overline{\beta} \, H_{\text{eff}})]$ and we chose $\overline{\beta}=\frac{1}{2}\left( \frac{1}{\TL} + \frac{1}{\TR}\right).$ Since $\rho_{\text{NESS}}$ must commute with the original Hamiltonian generating the dynamics, 
the effective Hamiltonian can only be a sum of conserved charges of the transverse field Ising model %commuting with $H$, i.e.,
\begin{equation}
H_{\text{eff}}= \sum_{n=0}^\infty \mu_n^{+} \, \mathcal{I}^+_n + \mu_n^{-} \, \mathcal{I}^-_{n} \,,
\end{equation} 
where the charges are given as \cite{FE13}
\begin{align}
& \mathcal{I}^+_n= \frac{i}{2}\sum_j a_{2j} (a_{2j+2n+1}+a_{2j-2n+1})-h \, a_{2j} (a_{2j+2n-1}+a_{2j-2n-1})\,, \\
& \mathcal{I}^-_{n-1}= -\frac{i}{2}\sum_j (a_{2j}a_{2j+2n} + a_{2j-1}a_{2j+2n-1})\,.
\end{align}
A state that is the exponential of a linear combination of conserved charges corresponding to a given Hamiltonian
%of such form 
is usually called a Generalized Gibbs Ensemble (GGE). 
%Let us also note that $\mathcal{I}^{+}_0$ is simply the transverse field Ising Ising Hamiltonian itself, while $\mathcal{I}^{-}_0$ is the spin-current operator. 
For the $h=1$ critical case,  one can immediately 
determine the coefficients $\mu_n^{\pm}$ from the  expectation values  \erf{aaNESS},
\begin{equation}
\mu_n^+= \delta_{n,0} \, , \; \; \;  \mu_n^{-}=\frac{16}{\pi \overline{\beta}} \left(\frac{1}{\TL} - \frac{1}{\TR} \right) \frac{n+1}{4n^2 +8n+3}\,,
\end{equation}
which implies that $H_{\text{eff}}$ decays algebraically, which also remains true
%and also 
in the $h \ne 1$ case.
% $H_{\text{eff}}$ decays algebraically.
%In particular, for the critical $h=1$ case, we obtain n -> n/2+1  
%\begin{equation}
%\mu_n^+= \delta_{n,1} \, , \;  .
%\end{equation}
%and also for $h \ne 1$ the
%which implies that the NESS is a . This is analogous to the case of
%of the transverse field Ising model
%which are very similar to the similar coefficients of the NESS belonging to the XX chain \cite{Ogata2002, EZ14}.
In summary, the NESS of the transverse field Ising model can be written in a GGE-like form but with a long ranged effective Hamiltonian. 
%From the form of the covariance matrix it is clear that by using 
%the 
%where the $\tilde{c}_k^\dagger$ and $\tilde{c}_k$ denotes the Fourier transforms of the $c^\dagger_j$ and $c^{\phantom{\dagger}}_j$ operators.
%
%
%
%Th
%\red{Let us note that the asymptotic state can be interpreted as the Gibbs state of an effective Hamiltonian
%\begin{equation}
%\rho_{NESS}= \frac{1}{Z} \exp (\overline{\beta}).
%\end{equation}
%As the NESS is a fermionic Gaussian state rom the covariance it is easy to check that the Gaussian state with above NESS correlation is equivalent to the following Hamiltonian 
%\begin{equation}
%\rho= \frac{1}{Z} 
%\end{equation}
%Fourier-transforming back, we obtain real space 
%\begin{equation}
%A
%\end{equation}
%}

\section{R\'enyi mutual information in the NESS\label{sec:RenyiMI}}

 In this section we show that in the NESS the R\'enyi mutual information of two adjacent intervals of length $L$ has logarithmic dependence on $L$ and present analytical results for the prefactor of the logarithm for indices $\alpha=1,2,3,4, 2^m$. The analytic expressions are compared with results obtained by numerical evaluation of the mutual information. Furthermore, we also study R\'enyi MI of higher index and study its dependence on the index and on the parameters of the system.

\subsection{Analytic results for the R\'enyi mutual information asymptotics}

From the results of Section~\ref{sec:NESS}, in particular from Eqs.~\eqref{aaNESS}, one can immediately determine the Majorana covariance matrix $\Gamma_{x,y}= \frac{i}{2}\langle [a_x, a_y] \rangle$ of the NESS. Due to translational invariance, the covariance matrix is a block-Toeplitz matrix, and thus can be expressed as the Fourier transform of a $2 \times 2$ matrix function called the \emph{symbol} of the block-Toeplitz matrix,
\begin{equation}
\label{NESSGamma}
\twomat{\Gamma_{2n-1, 2m-1}}{\Gamma_{2n-1,2m}  }
{\Gamma_{2n, 2m-1} }{\Gamma_{2n,2m}} =
\int_{-\pi}^{\pi} \frac{\dd k}{2\pi} \text{e}^{ik(m-n)} \Lambda(k)\,.
\end{equation}
For the transverse Ising NESS, the symbol is  given as
\begin{equation}
\label{NESSsymb}
\Lambda(k)=\twomat{i(\fR_{k}-\fL_{k})\, \mathrm{sgn}(k)}{ (\fL_{k}+\fR_{k}-1) \, e^{i\theta_{k}}}{ -(\fL_{k}+\fR_{k}-1) \, e^{-i\theta_{k}}}{i(\fR_{k}-\fL_{k})\, \mathrm{sgn}(k)}\,,
\end{equation}
where $\theta_k$ and $f^\text{R/L}_k$  are defined in Eqs. \erf{eq:theta} and \erf{eq:fq}, \erf{eq:eps}, respectively.
%where 
%\begin{align}
%&\tan \theta_k= \frac{\sin k}{h + \cos k} \, ,\; \; f^\text{L/R}(k) = \frac{1}{1{+}e^{\eps(k) /T_\text{L/R}}}, \; \, \text{with} \; \, \eps(k)= \sqrt{1+2h\cos k+h^2}.
%\end{align}
%
As the NESS is a Gaussian state, the von Neumann and R\'enyi entropies of a subsystem of $L$ consecutive spins can be calculated from the eigenvalues $\pm i \lambda^{(L)}_j$ of the $2L \times 2L$ reduced covariance matrix $\Gamma_{L}$ through the formula 
\begin{equation}
\label{Sformula}
S_L^{(\alpha)}=\sum_{j=1}^{L} s^{(\alpha)}\left(\lambda_j^{(L)}\right)\,,
\end{equation} 
where
\begin{align}
&s^{(\alpha)}(\lambda)= \frac{1}{1-\alpha}\log \left[ \left(\frac{1+ \lambda}{2}\right)^\alpha +  \left(\frac{1- \lambda}{2}\right)^\alpha  \right] \; \; \text{when} \; \; \alpha \ne 1\,,\\
&s^{(1)}(\lambda)= -\left(\frac{1+ \lambda}{2}\right) \log \left(\frac{1 + \lambda}{2}\right) - \left(\frac{1 - \lambda}{2}\right) \log \left(\frac{1- \lambda}{2}\right)\,.
\end{align}
This formula makes it possible to evaluate the entropy and mutual information numerically for large system sizes, and also an analytic treatment is possible through the use of  generalised Fisher--Hartwig formulas for the asymptotics of determinants of block-Toeplitz matrices.  Following Refs. \cite{JK04, KM05}, 
%one can connect the entropy to determinant asymptotic, 
one can use the residue theorem 
\begin{align}
S^{(\alpha)}_L = \sum_{j=1}^{L} s^{(\alpha)}(\lambda^{(L)}_{j})&= \frac{1}{2\pi i}\oint_{\mathcal{C}}  {\rm d} \lambda \,
s^{(\alpha)}(\lambda) \sum_{j=1}^L \frac{1}{\lambda - \lambda^{(L)}_{k}} \nonumber \\
&=\frac{1}{2\pi i}\oint_{\mathcal{C}} {\rm d}\lambda \,
s^{(\alpha)}(\lambda)\, \frac{{\rm d}\ln D_L(\lambda)}{ 2\, {\rm d}\lambda}\, ,
\label{contint}
\end{align}
where $D_L(\lambda)= \det(\lambda \, \unity -i \Gamma_L)$ and the contour $\mathcal{C}$ in the complex plane is encircling the real interval $[-1,1]$. In turn, also the von Neumann and R\'enyi mutual information between two adjacent blocks of size $L$ can be calculated as $I^{(\alpha)}_L=2S^{(\alpha)}_L-S^{(\alpha)}_{2L}$ using the above contour integration. 
A similar calculation was done for the NESS of the XX chain in Ref. \cite{EZ14}. %however, in that case the covariance matrix was of simple Toeplitz form. 
The big difference between the two cases is that instead of a simple Toeplitz matrix the covariance matrix is of block-Toeplitz type.
The rather lengthy calculation is delegated to Appendix A and we just state the results here. The R\'enyi mutual information can be shown to have a logarithmic asymptotics in the subsystem size, 
\be
I^{(\alpha)}_L=\sigma^{(\alpha)} \log L+ \text{const.}\,, \label{eq:asympt_alt}\\
\ee
and for $\alpha=1,2,3,4,2^m$ the prefactor of the logarithmic term can be explicitly obtained. Introducing the notations
\begin{subequations}
\label{eq:ab}
\begin{align}
\label{eq:ab1}
& a_1 =  \frac{1}{e^{-(1+h)/\TR}+1}\,, \; \; \qquad b_1\ =\frac{1}{e^{-(1+h)/\TL}+1}\,,\\
\label{eq:ab2}
&a_2=\frac{1}{e^{-(1-h)/\TR}+1}\,, \; \; \qquad b_2 =\frac{1}{e^{-(1-h)/\TL}+1}\,,
\end{align}  
\end{subequations}
and defining
\begin{align}  
\eta(w)=
\begin{cases}
\phantom{-}2\pi i \log (w) \; \; \text{when }  \arg(w) \in [0, \pi)\,,\\
-2\pi i \log(w) \; \; \text{when } \arg(w) \in [-\pi, 0)\,,
\end{cases}
\end{align}
they are given by
\begin{subequations}
\begin{align}
\sigma^{(1)}&=\frac{1}{2\pi^2} \sum_{i=1}^{2} \left[ a_i \, \Li{\frac{a_i{-}b_i}{a_i}} {+} (1{-}a_i) \Li{\frac{b_i{-}a_i}{1{-}a_i}}
  \nonumber \right. \\
& \left. \phantom{=\frac{1}{2\pi^2} \sum_{i=1}^{2}}  +  b_i \, \Li{\frac{b_i{-}a_i}{b_i}} {+}  (1{-}b_i) \Li{\frac{a_i{-}b_i}{1{-}b_i}}\right] \,, \label{eq:sigma1_alt}\\
\sigma^{(2)}&= -1-\frac{1}{2\pi^2} {\rm Re}\left[ \sum_{j=1}^2\log^2 \left(\frac{-2a_j + 1 +i}{2b_j -1-i}\right)  + 
\eta \left(\frac{2a_j - 1 -i}{2b_j -1-i}\right) \right] \label{eq:sigma2}\,,\\
\sigma^{(3)}&= -\frac{1}{2}-\frac{1}{4\pi^2} {\rm Re}\left[ \sum_{j=1}^2\log^2 \left(\frac{-2a_j + 1 +i/\sqrt{3}}{2b_j -1-i/\sqrt{3}}\right)  + 
\eta \left(\frac{2a_j - 1 -i/\sqrt{3}}{2b_j -1-i/\sqrt{3}}\right) \right]\,,\\
\sigma^{(4)}&={-}\frac{2}{3}-\frac{1}{6\pi^2} {\rm Re}\left[ \sum_{j=1}^2\log^2 \left(\frac{{-}2a_j {+} 1 {+} i \tan \frac{\pi}{8}}{2b_j {-}1{-}i \tan \frac{ \pi}{8}}\right)  + 
\eta \left(\frac{2a_j {-} 1 {-} i\tan \frac{\pi}{8}}{2b_j {-}1{-}i\tan \frac{\pi}{8}}\right) 
\right. \nonumber \\
&\left. \phantom{Re {-}2{-}\frac{1}{2\pi^2} \sum_{j=1}^2} \; \; + \log^2 \left(\frac{{-}2a_j {+} 1 {+}i \tan\frac{3\pi}{8}}{2b_j {-}1{-}i \tan \frac{3\pi }{8}}\right) + \eta \left(\frac{2a_j {-} 1 {-}i \tan \frac{3\pi}{8}}{2b_j {-}1{-}i \tan \frac{3\pi }{8}}\right)  \right]\,,\label{eq:sigma4}
\end{align}
%\sigma^{(2^m)}&= {-}\frac{2^{m-1}}{2^m{-}1} \nonumber \\
%-&\frac{1}{2\pi^2(2^m{-}1)} {\rm Re}\left[ \sum_{j=1}^2\sum_{k=1}^{2^{m-1}}\log^2 \left(\frac{{-}2a_j {+} 1 {+} i \tan \frac{(2k-1)\pi }{2^{m+1}}}{2b_j {-}1{-}i \tan\frac{(2k-1) \pi }{2^{m+1}}}\right)  + 
%\eta \left(\frac{2a_j {-} 1 {-} i \tan \frac{(2k-1)\pi}{2^{m+1}}}{2b_j {-}1{-}i \tan \frac{(2k-1)\pi}{2^{m+1}}}\right) \right].\\
\begin{align}
\sigma^{(2^m)}&= {-}\frac{2^{m-1}}{2^m{-}1} 
-\frac{1}{2\pi^2(2^m{-}1)} \sum_{k=1}^{2^{m-1}}\sum_{j=1}^2 {\rm Re}\left[  \log^2 \left(\frac{{-}2a_j {+} 1 {+} i \tan \frac{(2k-1)\pi }{2^{m+1}}}{2b_j {-}1{-}i \tan\frac{(2k-1) \pi }{2^{m+1}}}\right) \right. + \nonumber \\
&\phantom{= {-}\frac{2^{m-1}}{2^m{-}1} 
-\frac{1}{2\pi^2(2^m{-}1)} \sum_{k=1}^{2^{m-1}}\sum_{j=1}^2} \; \; \; \; \; \; \quad \left. \eta \left(\frac{2a_j {-} 1 {-} i \tan \frac{(2k-1)\pi}{2^{m+1}}}{2b_j {-}1{-}i \tan \frac{(2k-1)\pi}{2^{m+1}}}\right) \right]\label{eq:sigma2m}\,,
\end{align} 
where $\Li{w}$ denotes the dilogarithm function. 
Due to the structure of Eq.~\eqref{eq:sigma2m}, we can also get the analytical form of the R\'enyi 
%mutual information $I^{(\alpha)}_L$ 
prefactor in the $\alpha \to \infty$ limit:
\be
\sigma^{(\infty)}={-}\frac{1}{2}-\frac{1}{4\pi^3} \sum_{j=1}^2 \int_{-\frac{\pi}{2}}^{\frac{\pi}{2}} {\rm d} \theta \, \left[ \log^2 \left(\frac{{-}2a_j {+} 1 {+} i\tan \theta}{2b_j {-}1{-}i \tan \theta}\right) + \eta \left(\frac{2a_j {-} 1 {-}i \tan \theta}{2b_j {-}1{-}i\tan \theta}\right) \right]. \label{eq:sigmainf}
\ee
\label{sigmas}
\end{subequations}

%\begin{subequations}
%\begin{align}
%&\lim_{\alpha \to \infty} I^{\alpha}_L= \sigma^{(\infty)} \log L+ \text{const.} \, , \\
%&\sigma^{(\infty)}={-}\frac{1}{2}-\frac{1}{4\pi^3} \sum_{j=1}^2 \int_{-\frac{\pi}{2}}^{\frac{\pi}{2}} {\rm d} \theta \, \left[ \log^2 \left(\frac{{-}2a_j {+} 1 {+} i\tan \theta}{2b_j {-}1{-}i \tan \theta}\right) + \eta \left(\frac{2a_j {-} 1 {-}i \tan \theta}{2b_j {-}1{-}i\tan \theta}\right) \right]. \label{eq:sigmainf}
%\end{align}
%\label{sigmas}
%\end{subequations}

Let us make a few comments about these results. Equality of the initial temperatures, $T_\text{L}=T_\text{R},$ implies $a_j=b_j$  and the above formulas give $\sigma^{(\alpha)}=0$ for all $\alpha.$ Moreover, the expressions \erf{sigmas} are invariant under the transformation
\be
\label{duality}
(h,\TL,\TR)\longrightarrow (h^{-1},h/\TL,h/\TR)\,.
\ee
This is a simple manifestation of  the Kramers--Wannier duality, which in the Ising case can be regarded as a ``half-shift'' transformation, i.e. $a_x \to a_{x+1}$ (however, we emphasise that this transformation is non-local and does not strictly leave the R\'enyi MI  invariant, but the change can be only manifest in the subleading terms). Finally, let us note that for $\alpha>2$ the prefactor can be negative, this will be studied in detail in the next subsection.

\subsection{Negative R\'enyi mutual information}

\begin{figure}[t!]
\subfloat[$h=0.7,T_\text{L}=0.3,T_\text{R}=5$]{\includegraphics[width=.48\textwidth]{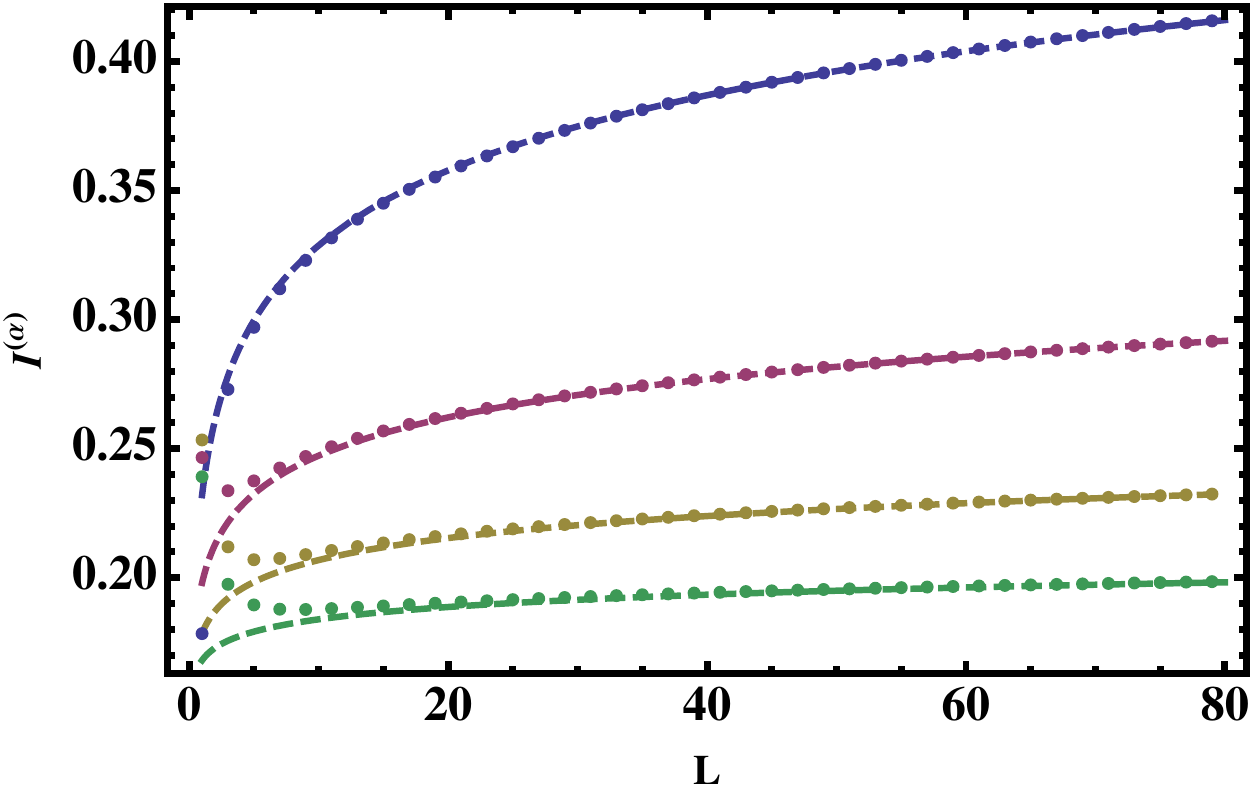}\label{fig:h07}}\hfill
\subfloat[$h=0.7,T_\text{L}=0.3,T_\text{R}=5$]{\includegraphics[width=.48\textwidth]{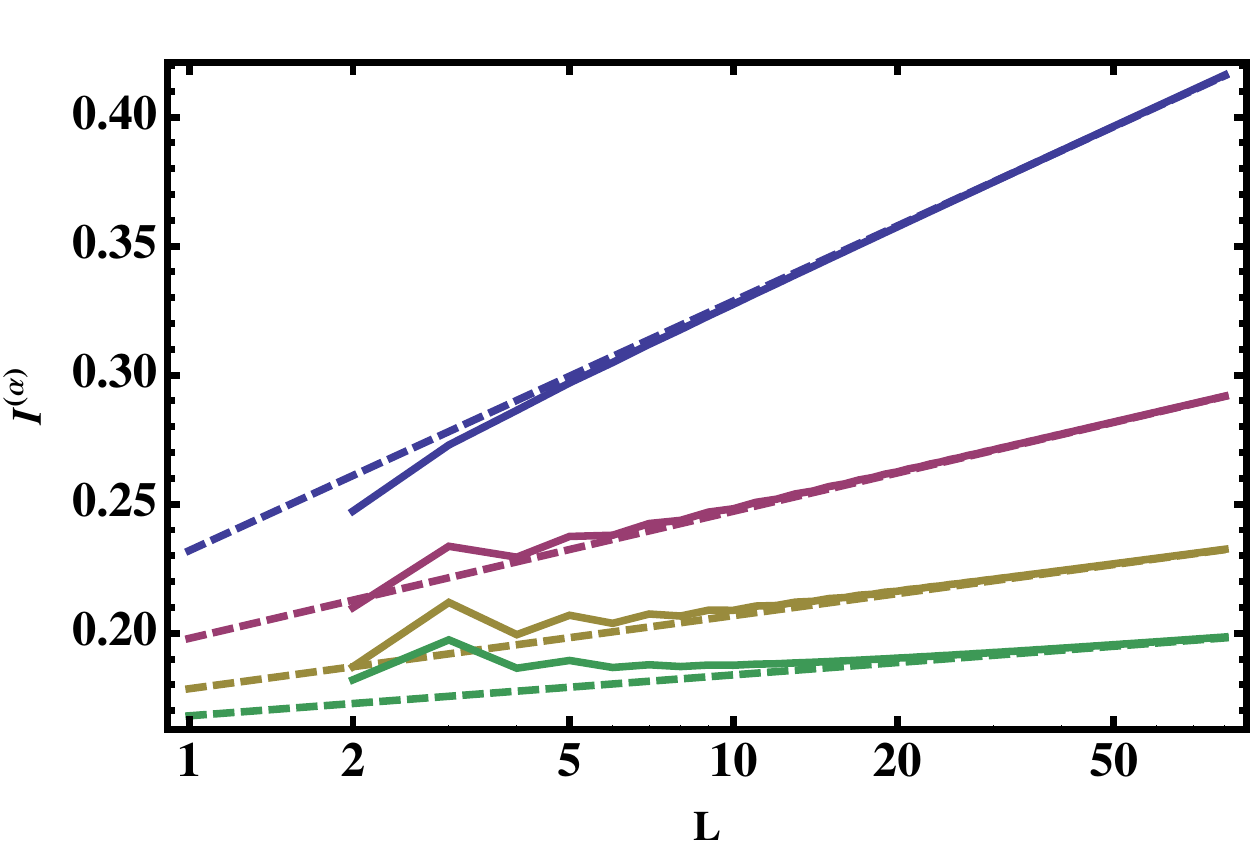}\label{fig:h17}}
\caption{\label{fig:MI1234} R\'enyi mutual information $I^{(\alpha)}$ as a function of the subsystem size $L$ for (from top to bottom) $\alpha=1,2,3,4$ at $h=0.7$ in the NESS with $T_\text{L}=0.3,T_\text{R}=5$ on (a) linear and (b) logarithmic scale. Numerical results are plotted in (a) dots and (b) continuous lines while the dashed lines show the function $\sigma^{(\alpha)} \log L + \text{const.}$, where the analytic results for $\sigma^{(\alpha)}$ given in Eqs.~\erf{eq:sigma1_alt}-\erf{eq:sigma4} and the constant is adjusted by hand.
}
\end{figure}

\begin{figure}[t!]
\subfloat[$h=0.1,T_\text{L}=0.1,T_\text{R}=1.5$]{\includegraphics[width=.47\textwidth]{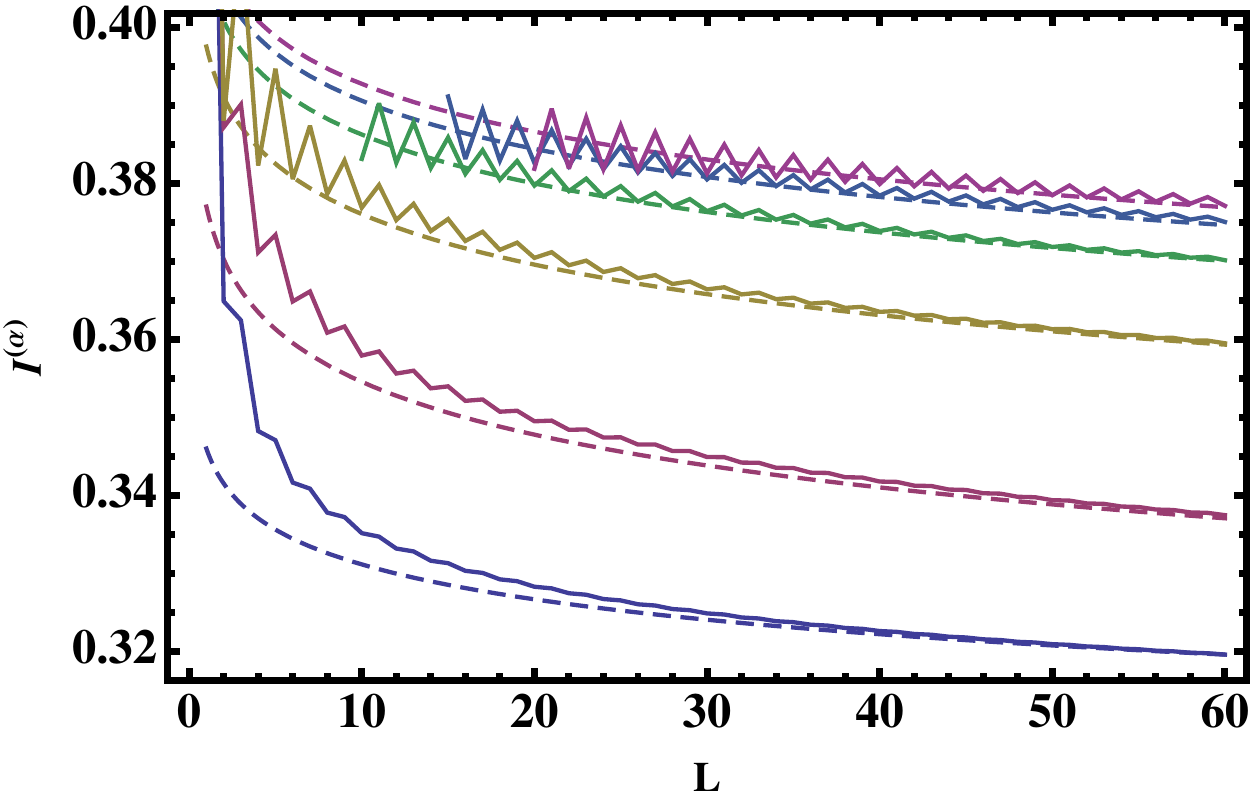}\label{fig:negferro}}\hfill
\subfloat[$h=10,T_\text{L}=1,T_\text{R}=15$]{\includegraphics[width=.49\textwidth]{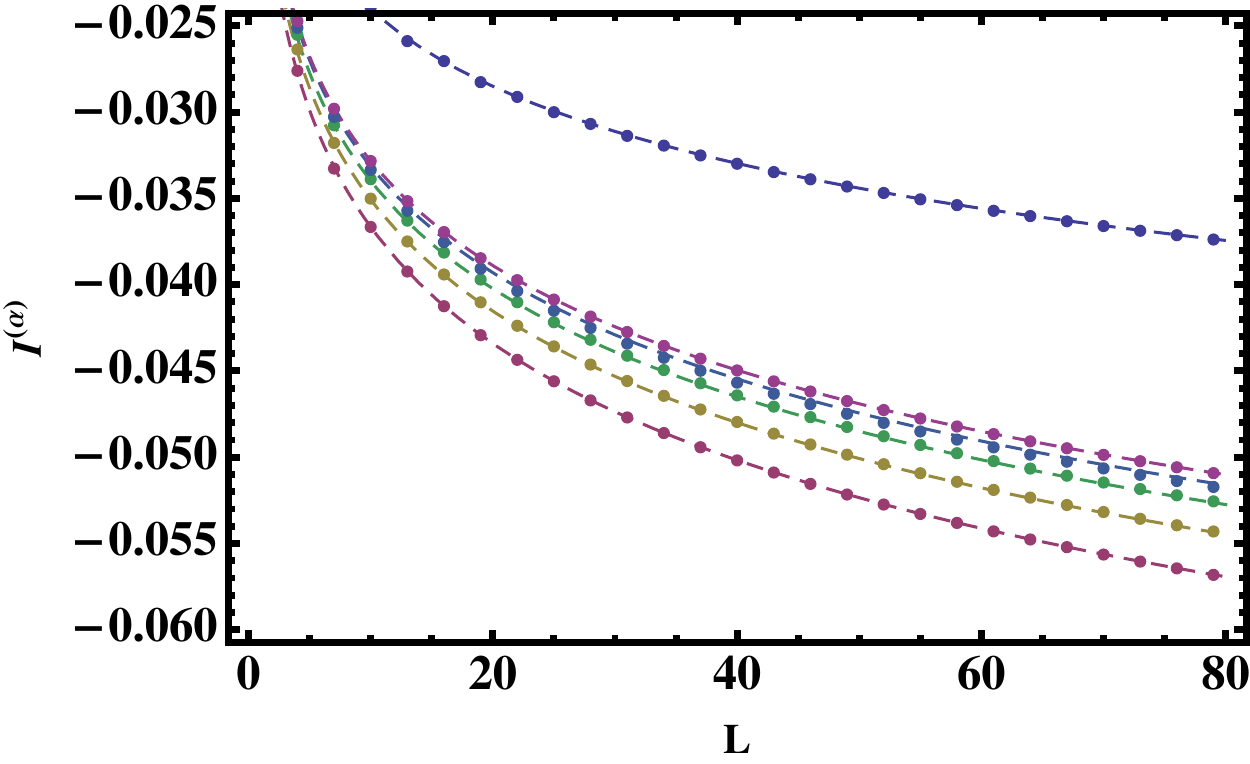}\label{fig:negpara}}
\caption{\label{fig:negMI} R\'enyi mutual information $I^{(\alpha)}$ as a function of the subsystem size $L$ in the NESS at (a) $h=0.1,$ $T_\text{L}=0.1,T_\text{R}=1.5$ for (from bottom to top) $\alpha=4,8,16,32,64,128;$ (b) $h=10,$ $T_\text{L}=1,T_\text{R}=15$ for (from bottom to top) $\alpha=8,16,32,64,128,4.$ Continuous lines (a) and dots (b) are numerical results and the dashed lines show the analytic result \erf{sigmas} with an adjusted additive constant.}
\end{figure}

We check our analytic expressions in Eq. \erf{sigmas} by comparing them to numerical results obtained by computing the sums \eqref{Sformula} over the eigenvalues of the covariance matrix \eqref{NESSGamma}. This is shown in Figs. \ref{fig:MI1234} and \ref{fig:negMI} where dots and continuous lines are numerical results and the dashed lines show the analytic expression, $I^{(\alpha)} = \sigma^{(\alpha)} \log L + \text{const.},$ with the constant shift adjusted by hand. We find excellent agreement in all cases.

%%%%%%%%%%%  FIGURES  %%%%%%%%%%%%%

\begin{figure}[t!]
\subfloat[$h=0.1,T_\text{L}=0.01,T_\text{R}=0.2$]{\includegraphics[width=.47\textwidth]{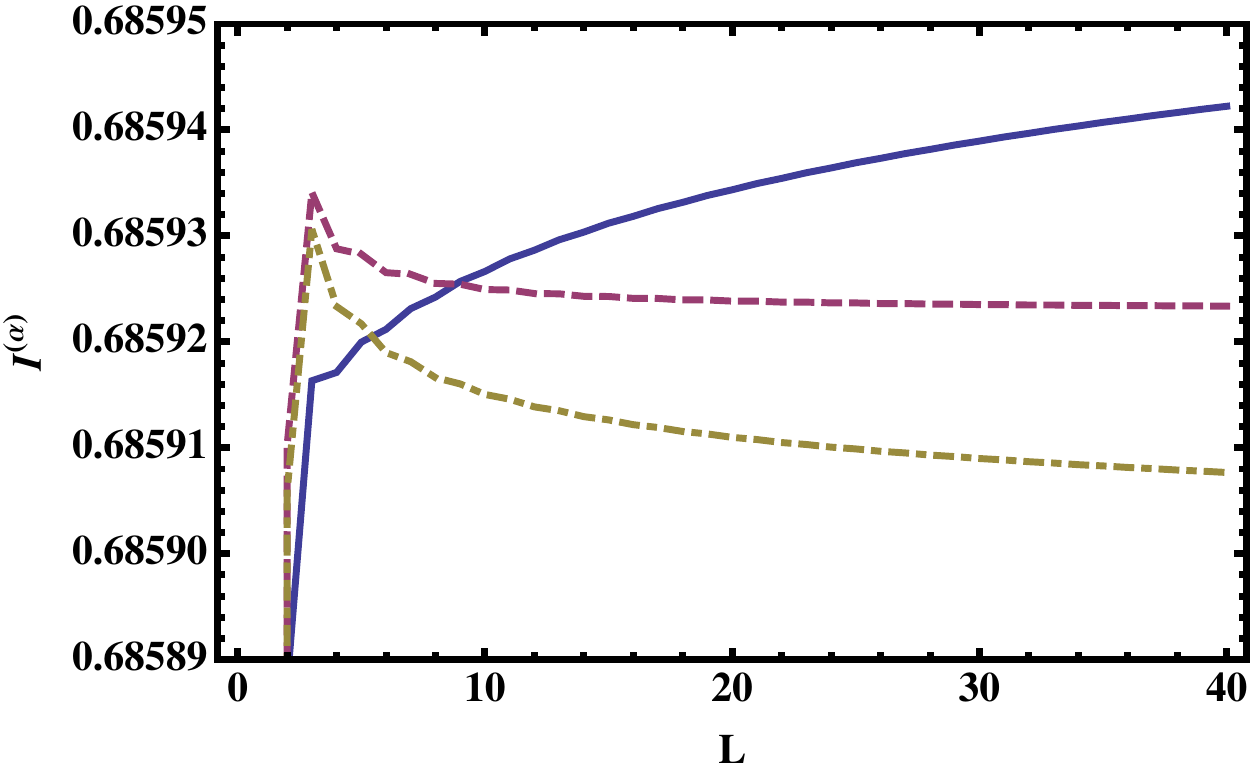}\label{fig:MI18-22}}
\hfill
\subfloat[$h=0.1,T_\text{L}=0.1,T_\text{R}=1.5$]{\includegraphics[width=.49\textwidth]{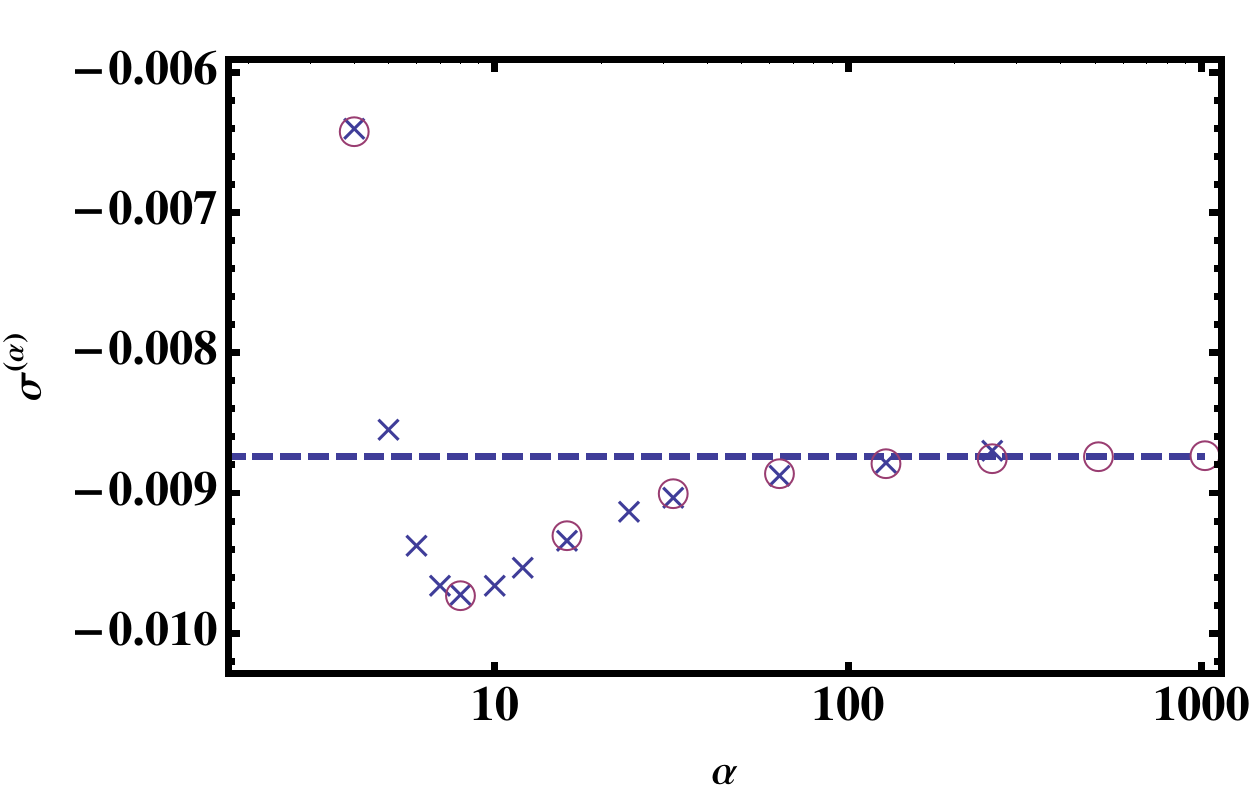}\label{fig:sigma(alpha)}}
\caption{(a) R\'enyi MI $I^{(\alpha)}$ as a function of $L$ for $\alpha=1.8,2,2.2$ (from top to bottom). (b) Dependence of $\sigma^{(\alpha)}$ on the R\'enyi index $\alpha.$ The analytic result \erf{eq:sigma2m} is shown in empty circles, numerical fits are shown in crosses, and the dashed line indicates $\sigma^{(\infty)}$ in \erf{eq:sigmainf}. 
}
\end{figure}

\begin{figure}[!h]
\subfloat[$h=0.1,T_\text{L}=0.1$]{\includegraphics[width=.48\textwidth]{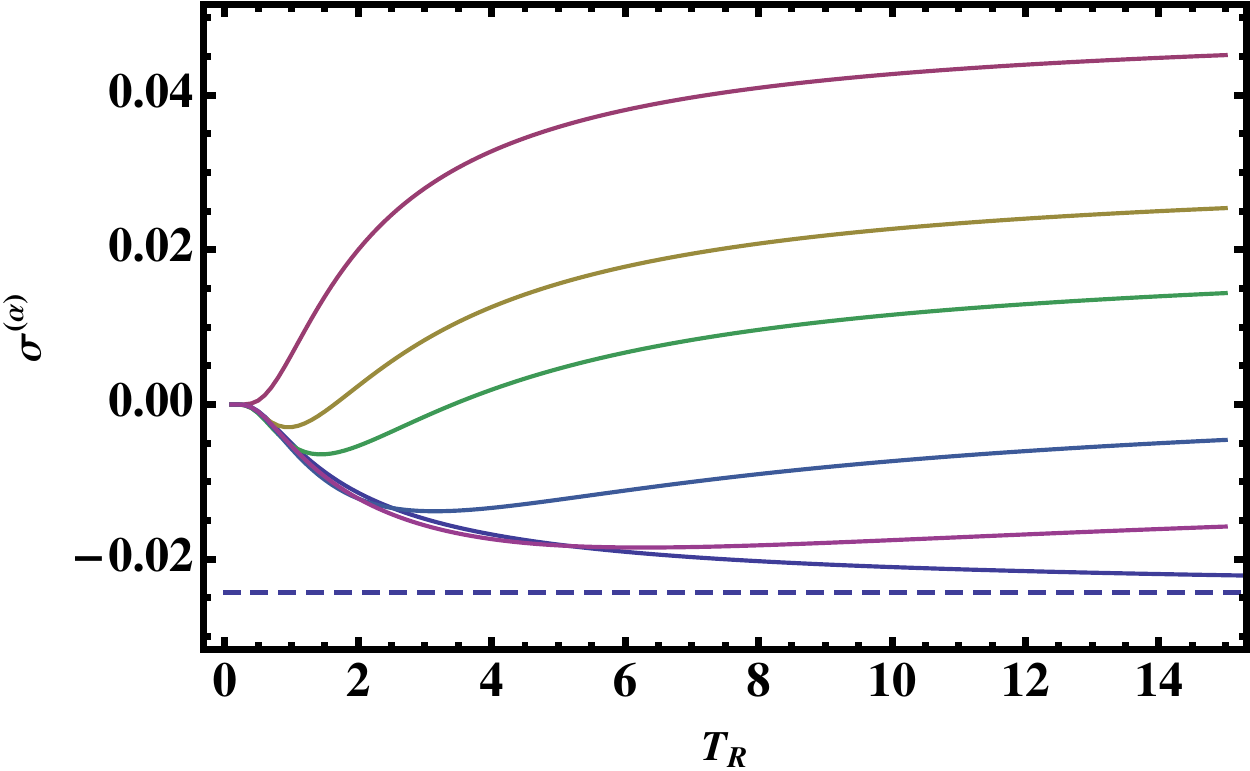}\label{fig:sigmaTR}}
\hfill
\subfloat[$T_\text{L}=0.1,T_\text{R}=4$]{\includegraphics[width=.48\textwidth]{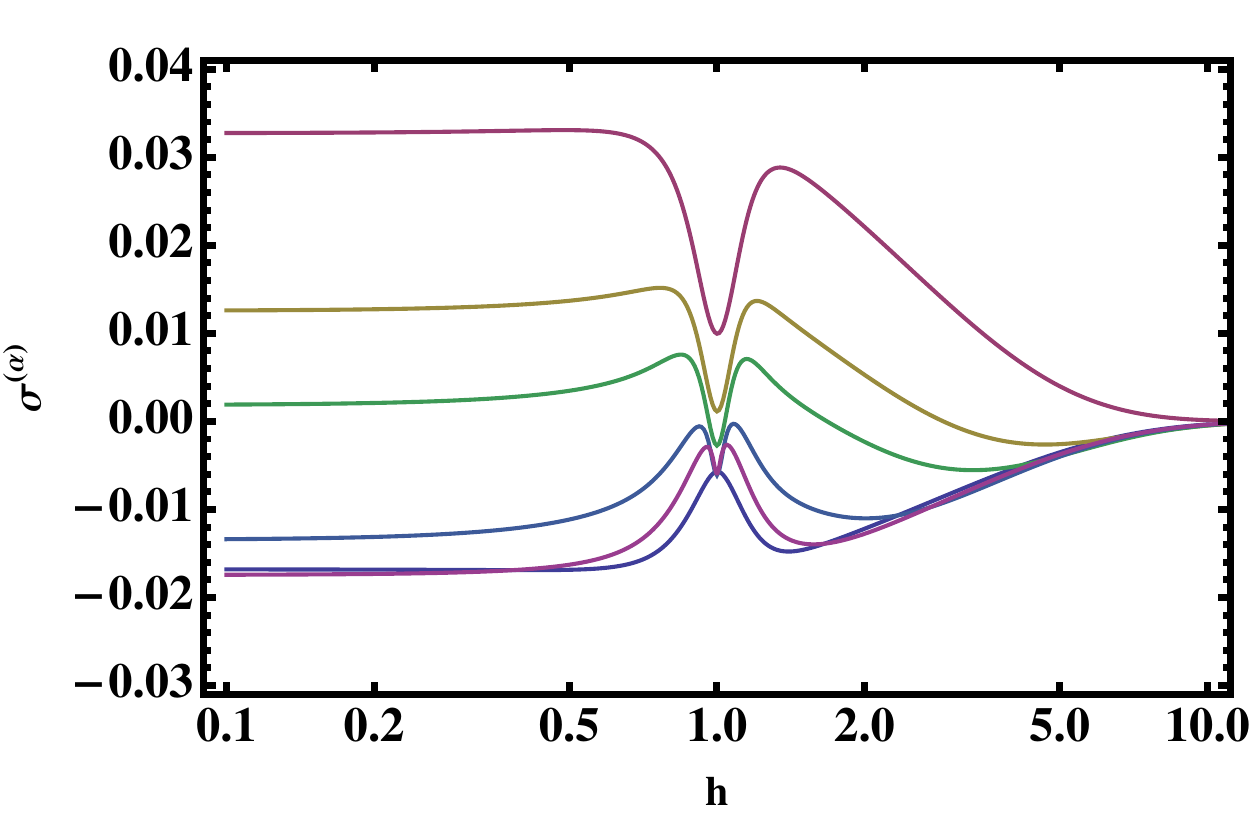}\label{fig:sigmah}}
\caption{ \label{fig:sigmadep}Prefactors $\sigma^{(\alpha)}$ with $\alpha=2,3,4,8,\infty$ (from top to bottom) (a) as a function of the right temperature $T_\text{R}$ for $h=0.1$ and $T_\text{L}=0.1;$ (b) as a function of $h$ for $T_\text{L}=0.1,$ $T_\text{R}=4$ on log-linear scale. In (a) the dashed line represents $\sigma^{(\infty)}$ at $T_\text{R}=\infty.$
}
\end{figure}

%%%%%%%%%  FIGURES  %%%%%%%%%%%

A surprising feature of the exact prefactors \erf{sigmas} is that for $\alpha\ge 3$  they can be negative implying that, rather counterintuitively, the corresponding R\'enyi MI monotonically {\em decreases} with the size $L$ of the subsystems for large enough 
$L.$ As a consequence, the R\'enyi MI can be {\em negative} and arbitrarily large in absolute value. This is shown in Fig. \ref{fig:negMI} both for a ferromagnetic and a paramagnetic case. The two cases are dual to each other under the transformation \erf{duality}, so the leading logarithmic contribution is the same but the constant shift and the subleading terms in general are different. For $h>1,$  the scaling form $I^{(\alpha)} = \sigma^{(\alpha)} \log L + \text{const.}$ is reached at smaller values of $L$ and the corrections to it are much smaller than for $h<1.$ In the latter case we also find an even-odd oscillating behaviour in $L$ which however decays as $L$ is increased. Fig. \ref{fig:negpara} shows an example where the R\'enyi MI is not only decreasing but it is also negative.

We have shown that $\sigma^{(3)}$ can take negative values, but it would be interesting to determine 
for which values of $\alpha$ can $\sigma^{(\alpha)}$ be negative. For non-integer R\'enyi index $\alpha$ we could only study this question numerically; and based on these investigations, we conjecture that for $\alpha>2$ there always exist parameters $h,T_\text{L},T_\text{R}$ such that $\sigma^{(\alpha)}<0,$ while for $\alpha<2$ the R\'enyi MI is always positive. As an illustration, in Fig. \ref{fig:MI18-22} we plot the R\'enyi MI for $\alpha=1.8,2,2.2$ at $h=0.1,T_\text{L}=0.01,T_\text{R}=0.2.$ For $\alpha=1.8$ $I^{(\alpha)}$ is increasing while for $\alpha=2.2$ it is decreasing. 

In both phases, $I^{(\alpha)}(L)$ converges to a limiting function as $\alpha\to\infty.$ Note that the approach is not monotonic, for example, in Fig. \ref{fig:negpara}, $I^{(8)}(L)<I^{(16)}(L)<\dots<I^{(\infty)}(L)<I^{(4)}(L)$ holds for the plotted range of $L.$ The prefactors $\sigma^{(\alpha)}$ are also non-monotonic in $\alpha$ as it is demonstrated in Fig. \ref{fig:sigma(alpha)}. Here the analytic results are plotted in circles while the crosses show the results of fitting the scaling form \erf{eq:asympt_alt} to the numerical data similar to those plotted in Figs. \ref{fig:MI1234} and \ref{fig:negMI}.

In Fig. \ref{fig:sigmadep} we study the dependence of $\sigma^{(\alpha)}$ on the temperatures and $h.$ As can be seen in Fig. \ref{fig:sigma(alpha)}, for fixed $h$ and $T_\text{L}$, the prefactor has a minimum as a function of $T_\text{R}$ for any finite $\alpha.$ The location of the minimum increases and the minimal value decreases with $\alpha,$ while $\sigma^{(\infty)}(T_\text{R})$ is a monotonically decreasing function approaching a limiting value as $T_\text{R}\to\infty.$ As a function of $h$ the prefactor has a local minimum at $h=1$ (see Fig. \ref{fig:sigmah}), however, the corresponding dip shrinks with increasing $\alpha$ and at $\alpha=\infty$ it completely disappears giving rise to a local maximum.

\subsection{Low temperature limit, comparison with CFT results}

\begin{figure}[t!]
{\includegraphics[width=.49\textwidth]{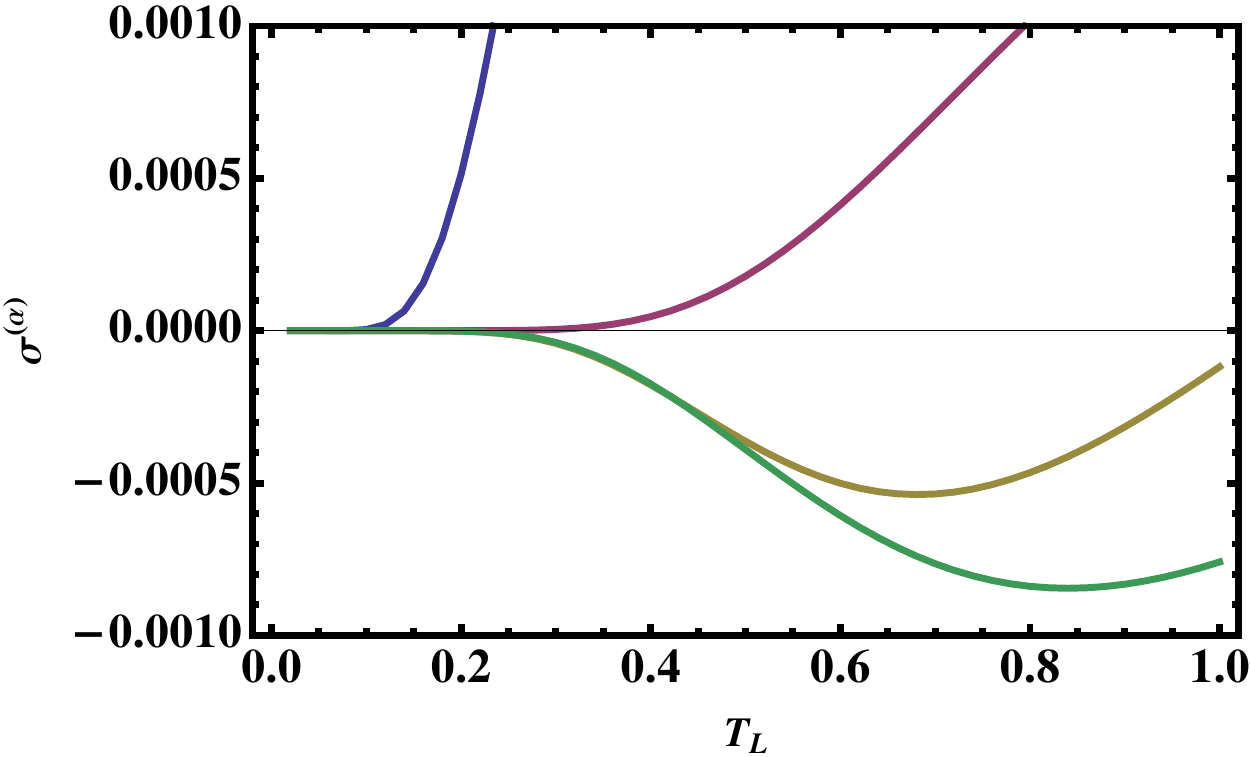}}\hfill
{\includegraphics[width=.45\textwidth]{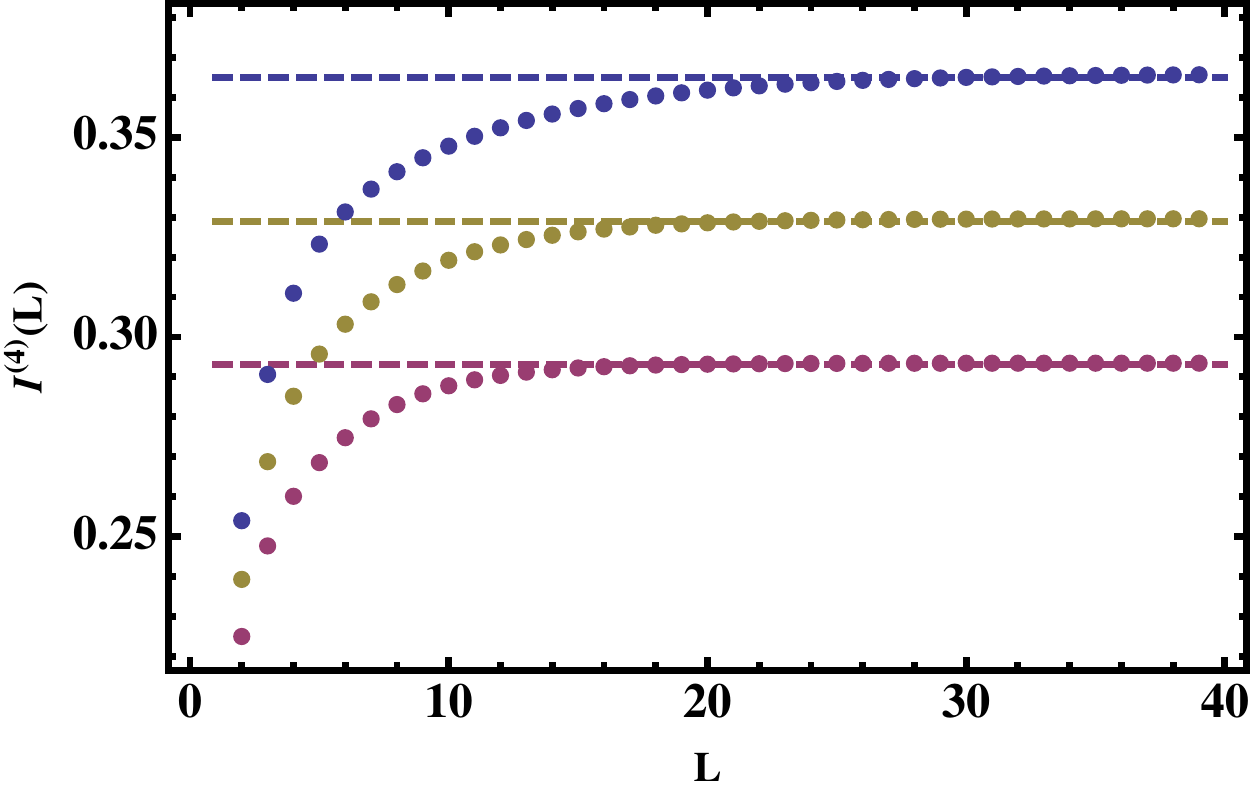}}
\caption{\label{fig:CFT} Low temperature behaviour of the R\'enyi mutual information at the critical point, $h=1$. (a) Temperature dependence of the prefactor $\sigma^{(\alpha)}$ for $\alpha=1,2,3,4$ (from top to bottom) for $T_\text{R}=2T_\text{L}.$ (b) R\'enyi mutual information  $I^{(4)}(L)$ as a function of the subsystem size $L$ in a thermal Gibbs state of $T=0.1$ (top) and $T=0.2$ (bottom), and in the NESS with $T_\text{L}=0.1,$ $T_\text{R}=0.2$ (middle). The dashed lines indicate the saturation values.
}
\end{figure}

We end the section by comparing our findings to conformal field theoretic results. The properties of non-equilibrium steady states have been extensively studied by CFT techniques (for a review, see Ref.~\cite{Bernard2016a}). In this line of research also the von Neumann and R\'enyi mutual information in the NESS was investigated \cite{Hoogeveen2014}. One of the central results in this context is that the $\alpha$-R\'enyi MI in the NESS generated from half-chains with temperature $\TL$ and $\TR$ is the average of the saturation value of the $\alpha$-R\'enyi MI for the respective Gibbs states 
\begin{equation}
I^{(\alpha)}_{\text{NESS}}(\TL, \TR)= \frac{1}{2}\left(I_{\text{Gibbs}}^{(\alpha)}(\TL) + I_{\text{Gibbs}}^{(\alpha)}(\TR)\right)\,. \label{eq:Rav}
\end{equation}
This seems to contradict our results, as for Gibbs states the R\'enyi MI is saturating, while for the NESS it is logarithmically diverging. However, we should bear in mind that CFT results are supposed to be valid in the low temperature limit. Indeed, we observed that for any $\alpha$ the prefactors of the logarithmic scaling tend to zero as $\TL, \TR \to 0$ (see some cases depicted in Fig.~\ref{fig:CFT}(a)), and the logarithmic scaling does not show up even for large subsystem sizes. Our numerical results suggest that for low temperatures (and for subsystem sizes where the logarithmic scaling is absent) the averaging property \eqref{eq:Rav} holds, see  Fig.~\ref{fig:CFT}(b) for an illustration. In this way the CFT results of Ref.~\cite{Hoogeveen2014} can be recovered for the Ising NESS.

\section{Time evolution of the R\'enyi mutual information \label{sec:MIevol}}

So far we have focused on the R\'enyi MI in the non-equilibrium steady state. Another and much more complicated aspect is the dynamics leading to the NESS. For the Ising model, the evolution of the von Neumann and R\'enyi entropies have already been studied for global \cite{Fagotti2008,Fagotti2010} and local quenches \cite{EKPP07,ISL09,Stephan2011}. We continue this line of investigations by asking how the R\'enyi MI evolves in time after joining the two halves of the system and how it reaches its stationary value. 

\begin{figure}[t!]
\subfloat[$\alpha=1$]{\includegraphics[width=.48\textwidth]{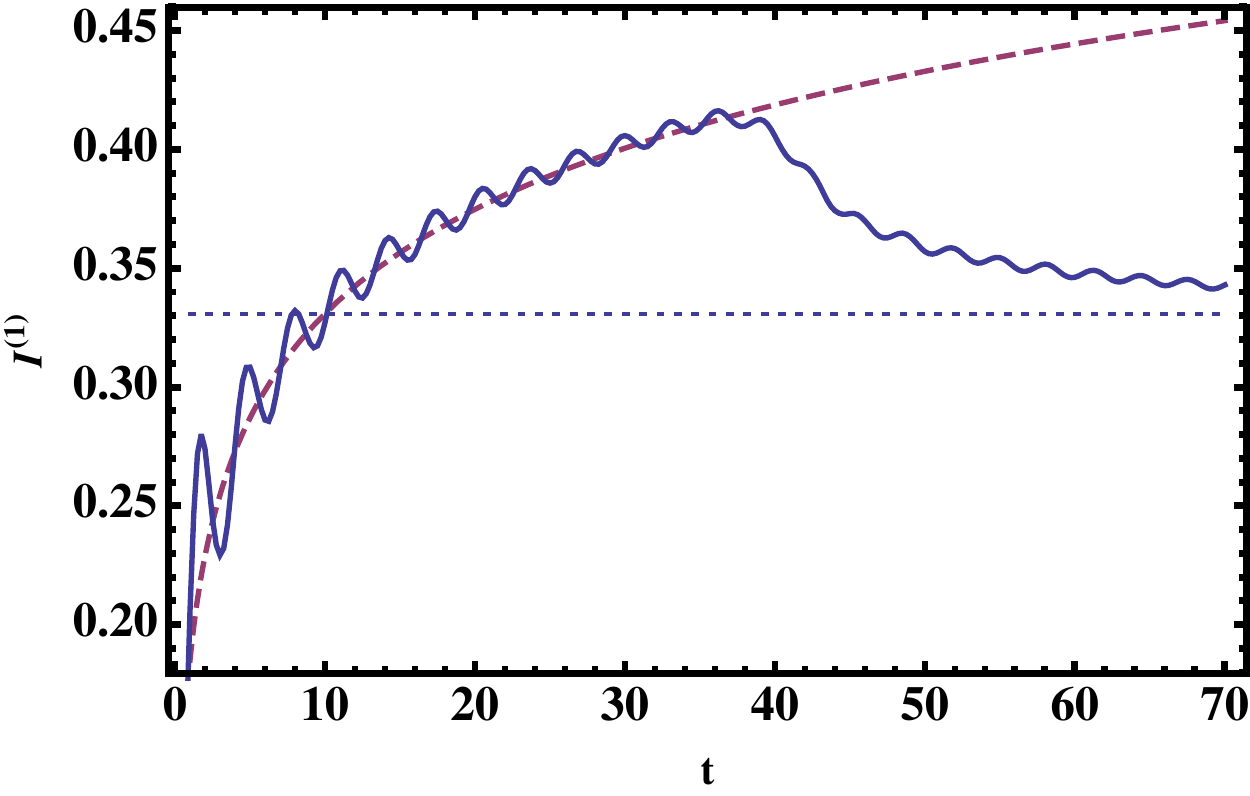}}
\hfill
\subfloat[$\alpha=2$]{\includegraphics[width=.48\textwidth]{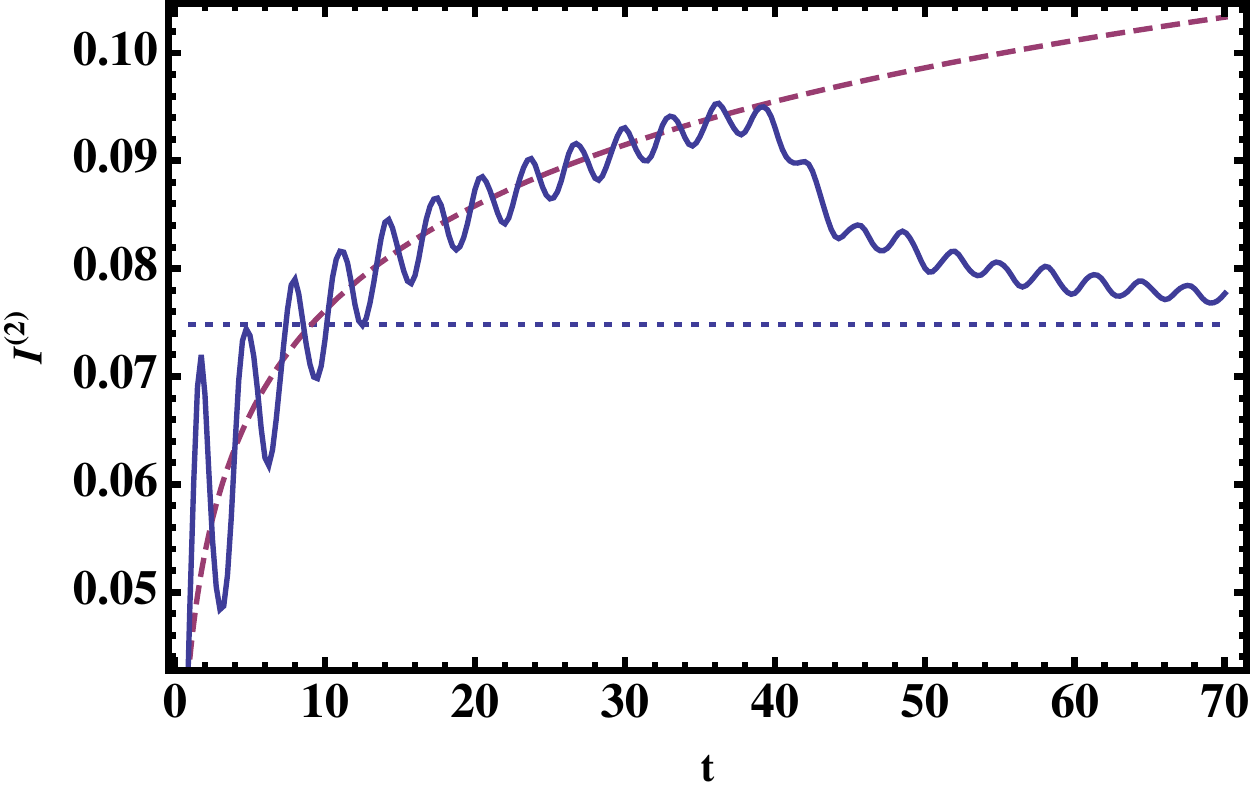}}\\
\subfloat[$\alpha=4$]{\includegraphics[width=.48\textwidth]{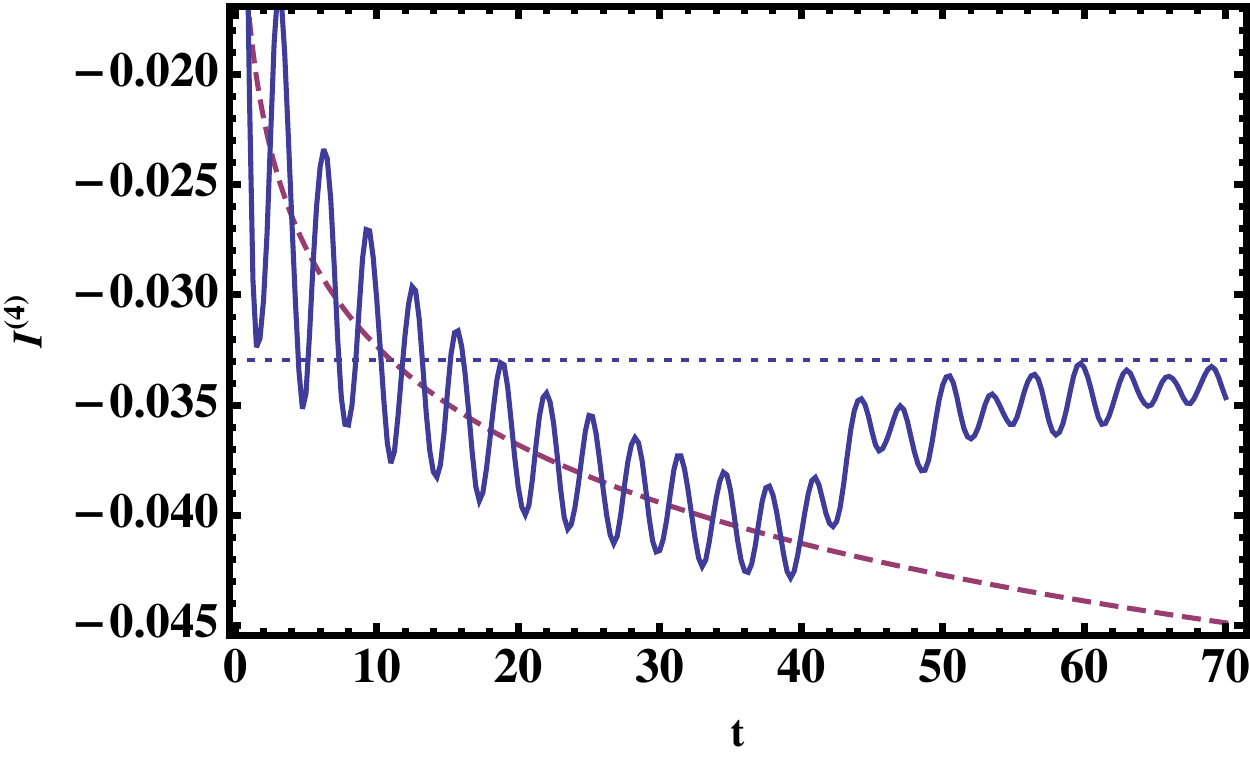}}
\hfill
\subfloat[$\alpha=\infty$]{\includegraphics[width=.48\textwidth]{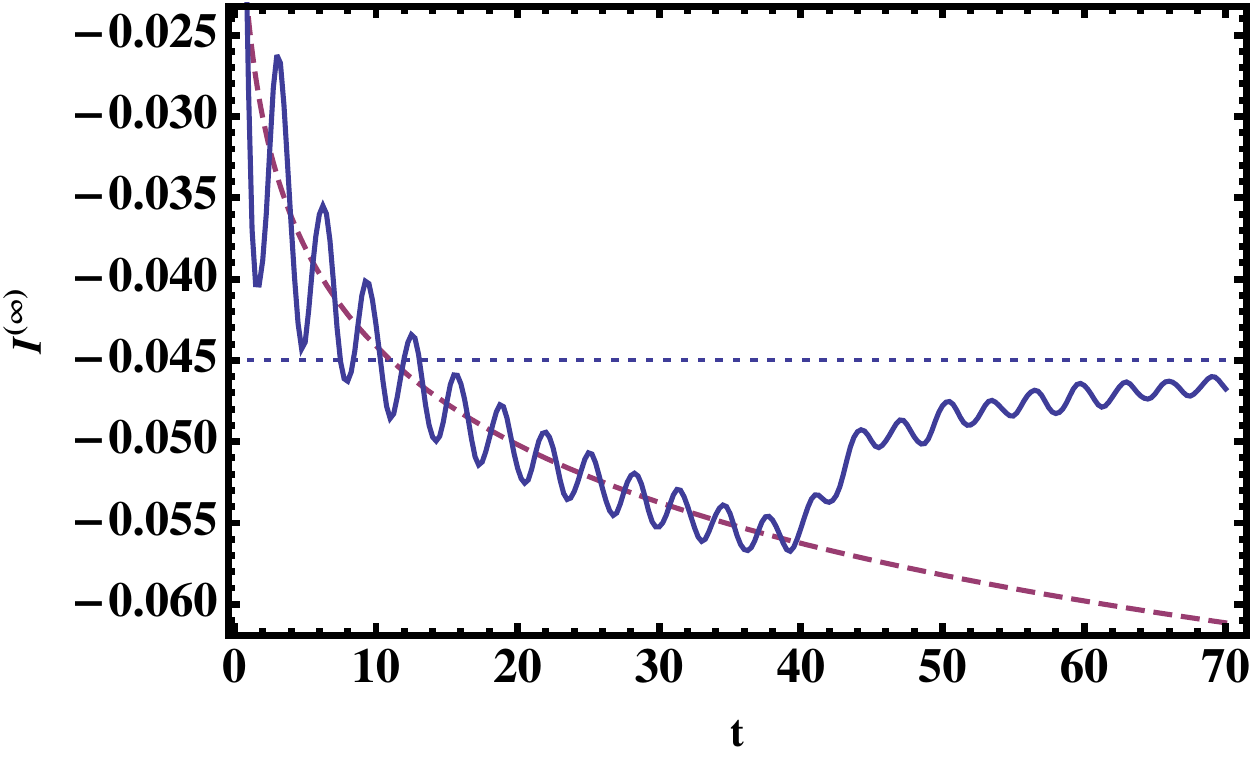}}
\caption{ \label{fig:MIth10} Time evolution of the R\'enyi MI for $\alpha=1,2,4,\infty$ at $h=10,T_\text{L}=1,T_\text{R}=15$ and $L=40.$ Numerical results are shown in solid line, the dashed lines correspond to $I^{(\alpha)}=\sigma^{(\alpha)}\log(t) + \text{const.}$ with the constants adjusted by hand. The horizontal dotted lines are the NESS result $I^{(\alpha)}=\sigma^{(\alpha)}\log(L) + \text{const.}$ where the constants were determined in Fig. \ref{fig:negpara}. }
\end{figure}

We do not attempt any analytic derivation  here  but resort to numerical investigations using time-evolution equations \eqref{ABcoeff} and \eqref{aacorr}. Some representative results in the paramagnetic phase are shown in Fig. \ref{fig:MIth10} for interval length $L=40.$ 
Note that there are cases when, quite oddly, the R\'enyi MI decreases after joining the two half chains.
We find that similarly to the XX model \cite{EZ14}, after an initial transient the MI evolves logarithmically in time up to $t\approx L:$
\be
I^{(\alpha)}(t) \approx \tilde \sigma^{(\alpha)} \log t + \text{const.}
\label{eq:It}
\ee
In our normalisation the maximal quasiparticle velocity is $v_\text{max} = \max (\ud \eps(k)/\ud k)=1$ for $h>1,$ so $t=L$ is the time necessary for the fastest quasiparticles to fly through and leave the interval. For $t>L$ there is a decay to the NESS value which is indicated by the dotted horizontal lines in Fig. \ref{fig:MIth10}. It is computed using the formula $I^{(\alpha)}_L = \sigma^{(\alpha)} \log L + \text{const.}$ with the analytic prefactors $\sigma^{(\alpha)}$ and the constant adjusted by hand as in Fig. \ref{fig:negpara}.

Based on our numerical findings we conjecture that the prefactor of $\log t$ is equal to the prefactor of the $\log L$ term in the NESS, that is,
\be
\tilde\sigma^{(\alpha)} = \sigma^{(\alpha)}\qquad\text{(conjecture)}\,.
\label{eq:sigmaeq}
\ee
Proving this equality is beyond the scope of our paper, but our numerical results strongly support this conjecture. In Fig. \ref{fig:MIth10} we plot in dashed line the function \erf{eq:It} using \erf{eq:sigmaeq} and our analytic results for $\sigma^{(\alpha)}.$ Similarly to the NESS fits, the constant is adjusted by hand. The agreement between the conjectured expression and the numerical results is excellent.

\begin{figure}[t!]
\subfloat[$\alpha=1$]{\includegraphics[width=.48\textwidth]{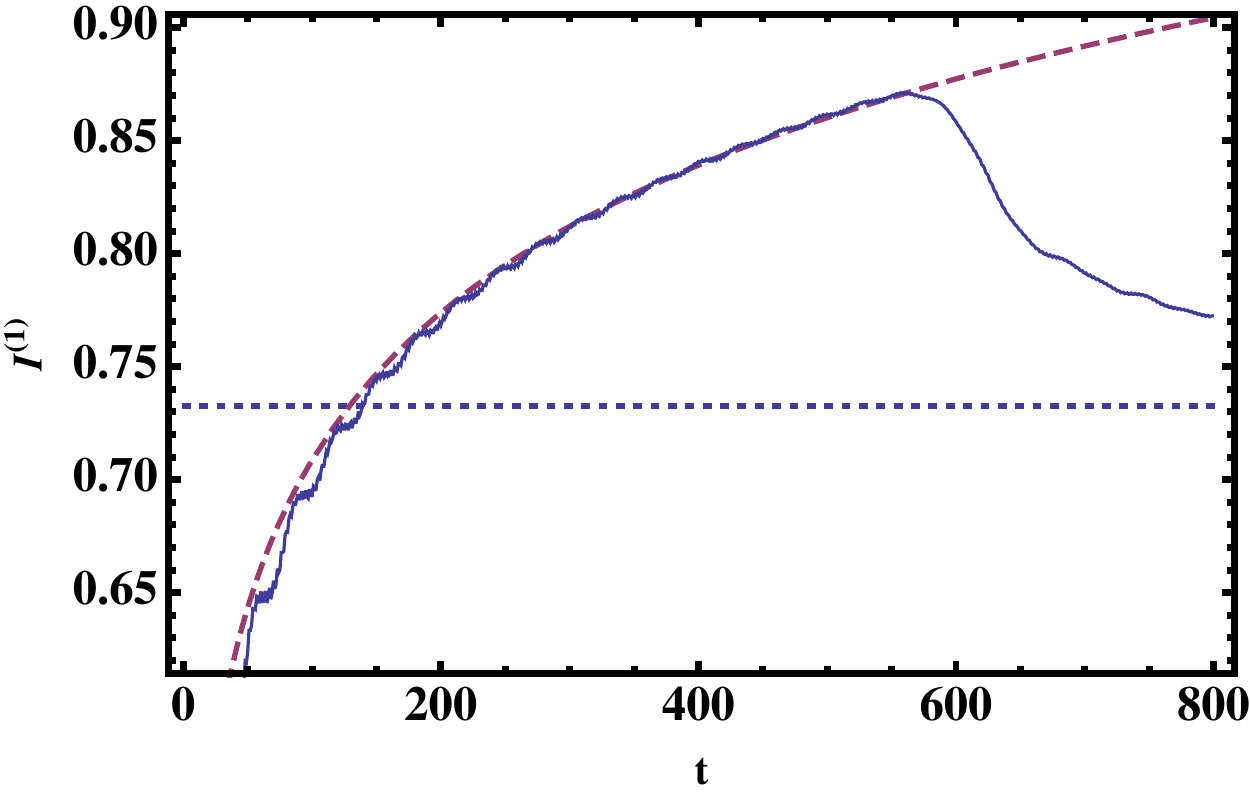}}
\hfill
\subfloat[$\alpha=2$]{\includegraphics[width=.48\textwidth]{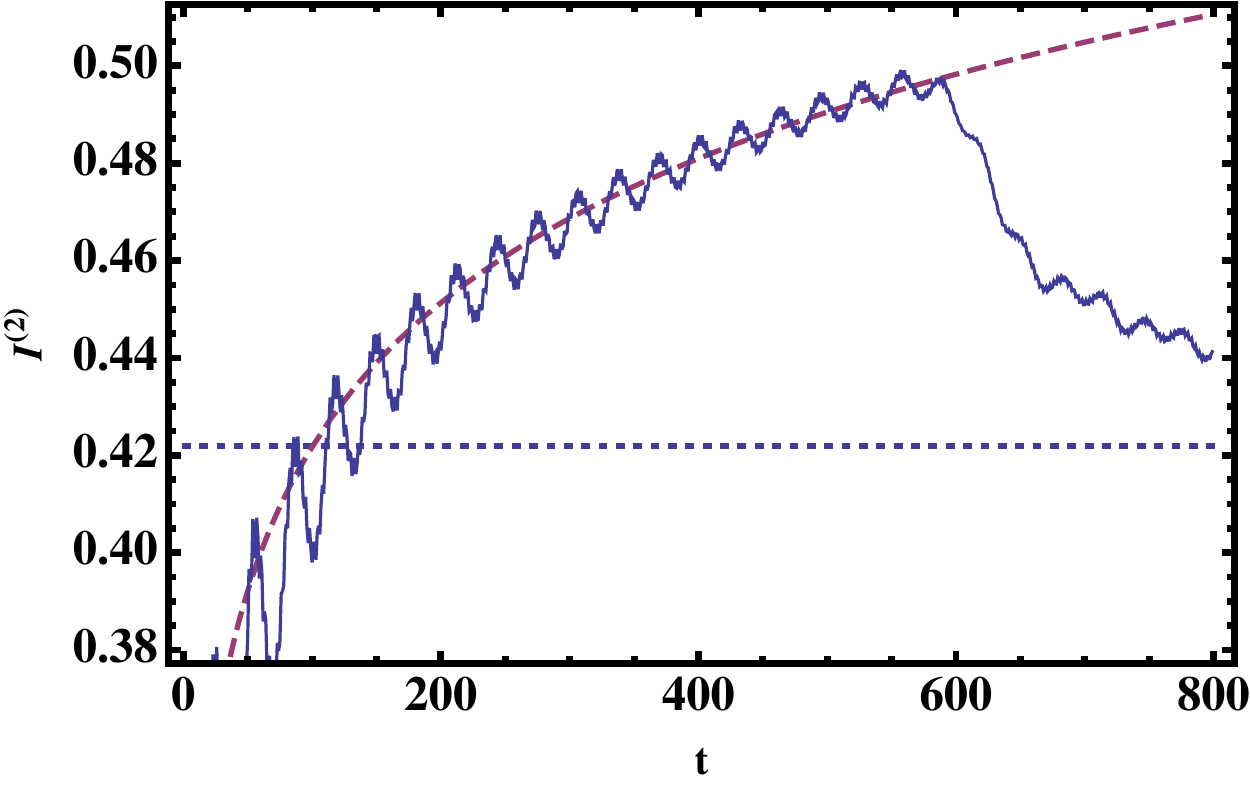}}\\
\subfloat[$\alpha=32$]{\includegraphics[width=.48\textwidth]{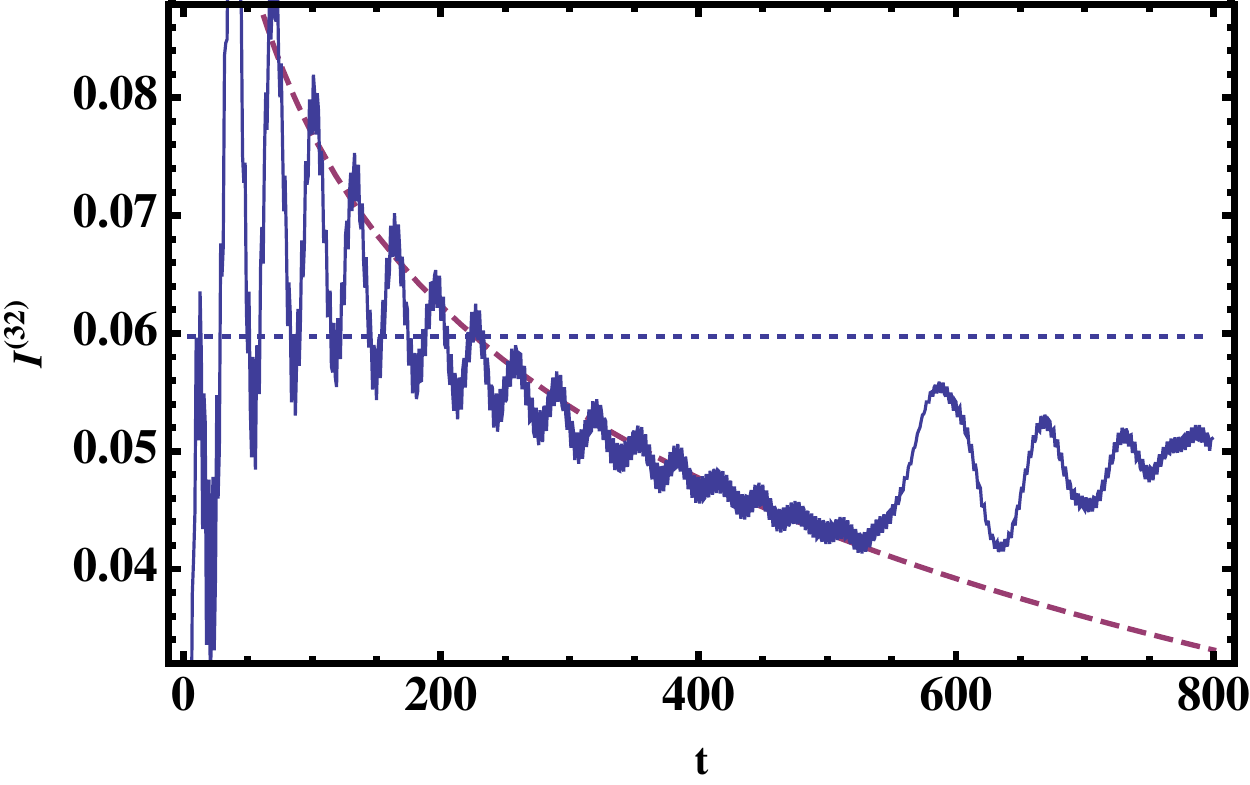}}
\hfill
\subfloat[$\alpha=\infty$]{\includegraphics[width=.48\textwidth]{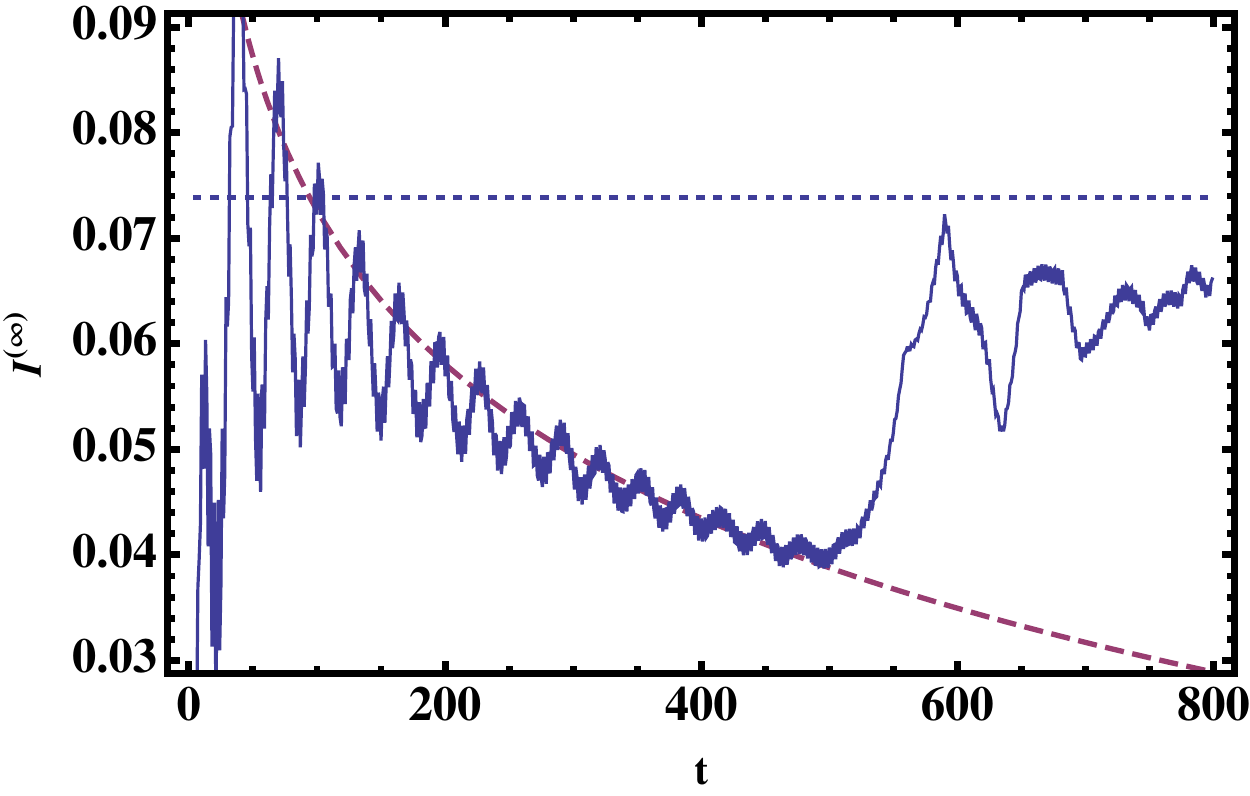}}
\caption{ \label{fig:MIth01} Time evolution of the R\'enyi MI for $\alpha=1,2,32,\infty$ at $h=0.1,T_\text{L}=0.1,T_\text{R}=10$ and $L=60.$ Numerical results are shown in solid line, the dashed line corresponds to $I^{(\alpha)}=\sigma^{(\alpha)}\log(t) + \text{const.}$ with the constants adjusted by hand. The horizontal dotted lines are the NESS result $I^{(\alpha)}=\sigma^{(\alpha)}\log(L) + \text{const.}$ where the constants were determined similarly to Fig. \ref{fig:negferro}. }
\end{figure}

In Fig. \ref{fig:MIth01} we present similar results in the ferromagnetic phase $h<1$ for $L=60.$ Here the amplitude of the oscillations are stronger than in the paramagnetic phase making the analysis more difficult. The agreement with Eqs. \erf{eq:It} and \erf{eq:sigmaeq} is still satisfactory.

Interestingly, for $h<1$ the deviation from the logarithmic behaviour \erf{eq:It} starts earlier than the quasiparticle picture would suggest. The maximal velocity for $h<1$ is $v_\text{max} = h,$ so the expected time where the logarithmic behaviour breaks down is $t=L/h.$ In Fig. \ref{fig:MIth01} we find, however, that for large R\'enyi index $\alpha$ Eq. \erf{eq:It} ceases to hold already for $t\approx500$ instead of $t=60/0.1=600.$ One is tempted to speculate that the higher index R\'enyi entropies may be more sensitive to the finite size of the quasiparticles. However, the physical reason behind this behaviour is presently unclear and deserves further study.

Decreasing $T_\text{R}$ while keeping $h$ and $T_\text{L}$ fixed the logarithmic behaviour \erf{eq:It} gradually disappears for $h<1$ (not shown here). We suspect that Eq. \erf{eq:It} still holds in an appropriate time window $1\ll t\ll L/h$ but this scaling regime is pushed towards larger times, which requires larger intervals that are more difficult to study numerically.

\section{Discussion and outlook \label{sec:outlook}}

In this work we studied the von Neumann and the R\'enyi mutual information between two touching intervals of length $L$ at the edges of two half infinite quantum Ising spin chains thermalised at different temperatures and subsequently glued together. Asymptotically a non-equilibrium steady state (NESS) is formed around the junction. We showed that in the NESS all the different types of MI depend logarithmically on the length of the intervals in the leading order, $I^{(\alpha)}=\sigma^{(\alpha)}\log(L)+\text{const.}$ We derived closed form exact analytic expressions \erf{sigmas} for the prefactor $\sigma^{(\alpha)}$ for $\alpha=2^m$ for all $m=0,1,2,\dots$ as well as for $\alpha=3$. We found that the dependence on $\alpha$ is not monotonic (c.f. Fig. \ref{fig:sigma(alpha)}). Taking $m$ to infinity allowed us to study the $\alpha\to\infty$ limit where we found a simple analytic expression \erf{eq:sigmainf} for $\sigma^{(\infty)}.$  We compared our analytic results to numerical calculations in finite systems.

Our most interesting finding is that the R\'enyi MI can assume negative values. We would like to stress that the setup featuring this behaviour is not a cooked-up or fine tuned one but rather a physical situation in a simple, paradigmatic system. The question whether this is related to the long-range nature of the effective Hamiltonian appearing in the GGE-like description of the final state deserves further study. We conjecture that for the Ising NESS $\alpha=2$ is a threshold value, that is $I^{(\alpha)}$ can be negative for $\alpha>2$ but it is always positive for $\alpha \le 2$ (see Fig. \ref{fig:MI18-22}). As the Ising NESS belongs to the family of fermionic Gaussian states, it would be interesting to investigate whether for this family the $2-$R\'enyi MI is always positive as was found for bosonic Gaussian states \cite{AGS12}.

We also studied the non-equilibrium time evolution of the R\'enyi MI after joining the two chains. It was shown that  in certain cases the R\'enyi MI can {\em decrease} in the course of the non-equilibrium time evolution.
Based on our numerical results, we conjecture that there is a time domain after the initial transient and before saturation takes place where the MI evolves logarithmically in time, $I^{(\alpha)}=\tilde\sigma^{(\alpha)}\log(t)+\text{const.}$ Moreover, we conjecture that the prefactor of the logarithmic term coincides with the prefactor of the logarithmic dependence on the subsystem size in the NESS, $\sigma^{(\alpha)}=\tilde\sigma^{(\alpha)}.$  The same behaviour was found in \cite{EZ14} for the XX spin chain. The proof of this statement is left for future work. 
Furthermore, we also made the observation that the logarithmic evolution of the MI stops earlier for higher R\'enyi indices than what a naive quasiparticle picture would suggest.

It is natural to expect that the negativity of the R\'enyi MI can also be observed in other systems and physical situations. The most probable candidates are current-carrying non-equilibrium steady states in other settings. It would also be interesting to study spin chains that cannot be mapped to free fermions, e.g., the integrable XXZ spin chain. 

Finally, it would be worthwhile to investigate in this context the alternative R\'enyi MI defined in terms of the R\'enyi divergences by Eq.~\eqref{eq:Reg_Renyi}. Studying these quantities would not only have natural consequences in the quantum information task of discriminating between many-body states (see, e.g. \cite{MHOF08}), but, presumably, would also yield a new tool for understanding many-body correlations.

\section*{Acknowledgment}
\noindent We would like to thank P.~Calabrese, V.~Eisler, M.~Mezei, and G.~Szirmai for useful discussions and valueble inputs. M.K. acknowledges funding from a ``Pr\'emium'' Postdoctoral Grant of the Hungarian Academy of Sciences and was partially supported by NKFIH grant no. K 119204. Z.Z. was supported by the DFG (CRC183, EI 519/9-1, EI 519/7-1) and the ERC (TAQ). Z.Z. would also like to thank the Simons Center for Geometry and Physics for hospitality, where some of the work has been carried out.

\section*{Appendix A: Analytical calculation of the R\'enyi mutual information asymptotics}

As stated in Section~\ref{sec:RenyiMI}, one can calculate the R\'enyi MI through the contour integral representation \eqref{contint} and by the use of a generalisation of the Fisher--Hartwig conjecture. Let us first recapitulate a simple version of the original conjecture. Consider for increasing $L$ Toeplitz matrices $T_L$ of dimension $L \times L$ defined by the symbol $\varphi(k)$, i.e.
\begin{equation}
(T_L)_{nm}= \int_{-\pi}^{\pi} \frac{\dd k}{2\pi} e^{ik(m-n)} \varphi(k) \, .
\end{equation}
Provided that $\phi(k)$ has the following factorisation form
\begin{equation}
\phi(k)=\psi(k) \prod_{r=1}^R v_{\gamma_r, \,q_r}(k) \, , 
\end{equation}
where $\psi(q)$ is a continuously  differentiable function and $v_{\gamma_r,\,q_r}$ describe
jumps at positions $k=q_r$ in the following form
\begin{equation}
v_{\gamma_r,\,q_r}(k)=\exp [-i\gamma_r (\pi{-}k{+}q_r)]\,, \qquad q_r<k <2\pi{+}q_r\,, 
\end{equation}
then the $L \to \infty$ asymptotics of the determinant is
\begin{equation} 
\det(T_L)= \left({\cal F}[\psi]\right)^{\mathrm{L}}
\left(\prod_{r=1}^R {\mathrm{L}}^{-\gamma_r^2}\right)
{\cal E}[\psi, \{\gamma_r\},\{q_r\}]\,, \label{fh}
\end{equation}
where ${\cal F}[\psi]=\exp \left(\frac{1}{2\pi} \int_0^{2\pi}\ln
\psi(k) \mathrm{d} k \right)$, and the ${\cal E}$ term does not depend on $L$. 

For translation invariant Gaussian states, the covariance matrix is usually not a simple Toeplitz matrix,
rather a block-Toeplitz matrix (composed of $2 \times 2$ blocks).  For the case when the covariance matrix factorises in the form 
\begin{equation} \label{eq:fact}
\Gamma_L=\twomat{0}{1}{-1}{0} \otimes T_L
\end{equation}
with $T_L$ a Toeplitz matrix, the asymptotics of $\det \Gamma_L$ can be derived from the Fisher--Hartwig conjecture. For such models, using the contour integral representation, it can be shown that the ${\cal F}[\psi]^L$ factor will give rise to the linear term in the entropy asymptotics (which is zero for pure Gaussian states), while a logarithmic subleading term is induced by the $\prod_{r=1}^R {\mathrm{L}}^{-\gamma_r^2}$ factor, and the final ${\cal E}[\psi, \{\gamma_r\},\{q_r\}]$ factor only provides a constant term \cite{JK04,KM05,KZ10, CalabreseEssler,FEFS14}. Let us also mention that when calculating the mutual information, the linear term drops out and the logarithmic term provides the leading order in the asymptotics \cite{EZ14}. 

%An example where the above shown Toeplitz scenario can be is the 

The covariance matrix of the Ising NESS cannot be factorised in the form of Eq.~\eqref{eq:fact}, thus one has to use generalisations of the Fisher--Hartwig conjecture for obtaining the mutual information asymptotics. The Fisher--Hartwig method presented above has been extended in many directions \cite{IJK05, EC05, FIK08}, and also the original conjecture has been strengthened. 
%\cite{FEFQ15}. 
%through a block Toeplitz matrix that cannot be factorised as \ref{eq:fact} corresponding to the block Toeplitz symbol $\Lambda$. There are many generalised versions of the Fisher--Hartwig conjecture.
In particular, in Ref. \cite{FEFQ15} a generalisation was proposed for the case of a block symbol $\Lambda(q)$ that is continuously differentiable apart from a finite number of points $k=q_r$ $(r=1 \dots R)$ where it has jumps satisfying the condition $\lim_{\epsilon \to 0} [\Lambda(q_r-\epsilon), \Lambda(q_r+\epsilon)]=0$. Such $2 \times 2$ block symbols can be written as
\begin{equation}
\Lambda(k)= U^\dagger (k) \left( \Psi(k) \prod_{r=1}^{R} V_r(k) \right) U(k)\,,
\end{equation}
where $\Psi(k)$ is continuously differentiable diagonal $2\times 2$ matrix symbol,  $U(k)$ is a continuously differentiable function of unitary matrices that diagonalise $\Lambda(k)$, and the jump matrices $V_r(k)$ are of the form
\begin{equation}
V_r(k)=\twomat{\exp [-i\gamma_r (\pi{-}k{+}q_r)]}{0}{0}{\exp [-i\delta_r (\pi{-}k{+}q_r)]}\,,  \qquad  q_r<k <2\pi{+}q_r \, . 
\end{equation}
According to the generalised Fisher--Hartwig conjecture, the determinant  can again be factorised in a similar form than that of Eq.~\eqref{fh}, i.e.  $ \det(\Gamma_L)=\mathcal{F}^L (\prod_{r=1}^R {\mathrm{L}}^{-\gamma^{2}_r-\delta^{2}_r}){\cal E}$, where $\mathcal{F}$ and $\mathcal{E}$ do not depend on  $L$. In the block-Toeplitz case the explicit form of $\mathcal{F}$ and $\mathcal{E}$ is not known. However, since the linear term of the entropy (corresponding to $\mathcal{F}^L$) drops out from the mutual information asymptotics, we are able to calculate the leading order correction of $I^{(\alpha)}_L$.

Let us turn attention to the particular case of the Ising NESS. As discussed earlier, in order to calculate the von Neumann and R\'enyi entropies using the contour integral \eqref{contint}, we have to consider the Toeplitz matrix corresponding to the symbol $\lambda \unity -i\Lambda(k)$.
There are two jumps in this symbol, at $k=0$ and at $k=\pi/2$. The diagonal elements of the jump matrices $V_1$ and $V_2$  are the following:
\begin{subequations}
\begin{align}
& \gamma_1(\lambda){=}\frac{1}{2\pi i} \log \left( \frac{\lambda -2(e^{-(h+1)/\TR}{+}1)^{-1}+1}{\lambda-2(e^{-(h+1)/\TL}{+}1)^{-1}+1}\right)\,,\\
& \delta_1(\lambda){=} \frac{1}{2\pi i} \log \left( \frac{\lambda -2(e^{-(h-1)/\TR}{+}1)^{-1}+1}{\lambda-2(e^{-(h-1)/\TL}{+}1)^{-1}+1}\right)\,,\\
& \gamma_2 (\lambda){=} \frac{1}{2\pi i} \log \left( \frac{\lambda -2(e^{(h+1)/\TR}+1)^{-1}+1}{\lambda- 2(e^{(h+1)/\TL}+1)^{-1}+1}\right)\,,\\
& \delta_2 (\lambda){=} \frac{1}{2\pi i} \log \left( \frac{\lambda -2(e^{(h-1)/\TR}+1)^{-1}+1}{\lambda- 2(e^{(h-1)/\TL}+1)^{-1}+1}\right)\,.
\end{align}
\end{subequations}
Taking the logarithm of  $D_L(\lambda)= \det(\lambda \, \unity -i \Gamma_L),$
\begin{equation}
\log D_L(\lambda)=  L \log \mathcal{F(\lambda)}   - (\gamma^2_1(\lambda) + \delta^2_1(\lambda) + \gamma^2_2(\lambda) + \delta^2_2(\lambda))
\log L + \log \mathcal{E}(\lambda) \,.
\end{equation}
We will drop the $\log \mathcal{E}(\lambda)$ term, as it only gives an $L$-independent value and we calculate 
$I^{(\alpha)}_L$ up to $\mathcal{O}(1)$ in $L$.
Taking the derivative of $\log D_L $, we obtain:
\begin{align}
&\frac{\mathrm{d} \log D_L(\lambda)} {\mathrm{d} \lambda} = 
\frac{\mathrm{d} \log (\mathcal{F}(\lambda))} {\mathrm{d} \lambda}  L \, + \nonumber \\
&\; \; \; \left[\frac{2(b_1{-}a_1)}{ \pi i}\left( \frac{\gamma_1(\lambda)}{(2a_1{-}1{-}\lambda)(2b_1{-}1{-}\lambda)} {+} \frac{\gamma_2(\lambda)}{(1{-}2a_1{-}\lambda)(1{-}2b_1{-}\lambda)}  \right) +\right.   \nonumber\\
&\; \; \; \, \left. \frac{2(b_2{-}a_2)}{ \pi i}\left( \frac{\delta_1(\lambda)}{(2a_2{-}1{-}\lambda)(2b_2{-}1{-}\lambda)} + \frac{\delta_2(\lambda)}{(1{-}2a_2{-}\lambda)(1{-}2b_2{-}\lambda)}  \right)\right]\log L \,,
\end{align}
where $a_1, a_2, b_1, b_2$ are defined in Eq.~\eqref{eq:ab}.
%\begin{align}
%%\label{eq:ab}
%& a_1 =  \frac{1}{e^{-\betaR(1+h)}+1}\,, \; \; \; b_1\ =\frac{1}{e^{-\betaL(1+h)}+1}\,,\\
%&a_2=\frac{1}{e^{-\betaR(1-h)}+1}\,, \; \; \; b_2 =\frac{1}{e^{-\betaL(1-h)}+1}\,. 
%\end{align} 

When calculating the mutual information the term proportional to $L$ drops out, thus, using Eq.~\eqref{contint},  we obtain that 
\begin{align}
I^{(\alpha)}_L&= \frac{a_1-b_1}{2\pi^2} \left(\oint_{\mathcal{C}_1} \mathrm{d} \lambda
\frac{s^{(\alpha)}(\lambda) \, 
\gamma_1(\lambda)}{(2a_1{-}1{-} \lambda)(2b_1{-}1{-}\lambda)} + 
\oint_{\mathcal{C}_2} \mathrm{d} \lambda\frac{s^{(\alpha)}(\lambda) \, 
\gamma_2(\lambda)}{(1{-}2a_1{-} \lambda)(1{-}2b_1{-}\lambda)}\right) \log L \nonumber \\
&+\frac{a_2-b_2}{2\pi^2} \left(\oint_{\mathcal{D}_1} \mathrm{d} \lambda
\frac{s^{(\alpha)}(\lambda) \, 
\delta_1(\lambda)}{(2a_2{-}1{-} \lambda)(b_2{-}1{-}\lambda)} + 
\oint_{\mathcal{D}_2} \mathrm{d} \lambda \frac{s^{(\alpha)}(\lambda) \, 
\delta_2(\lambda)}{(1{-}2a_2{-} \lambda)(1{-}2b_2{-}\lambda)}\right) \log L\,, \label{eq:4cont}
\end{align}
where, due to the position of the divergences and cuts of the integration kernel, the contour $\mathcal{C}$ encircling the interval $[-1,1]$ could be broken up to four smaller contours $\mathcal{C}_1,$ $\mathcal{C}_2,$ $\mathcal{D}_1,$ and $\mathcal{D}_2$ which encircle the branch cuts of the four denominators. For example, contour $\mathcal{C}_1$ encircles the interval\footnote{Here and below we assume $b_1>a_1$ but the calculation is analogous in all other cases.} $[2a_1-1,2b_1-1].$ 
%with ${k=1,\ldots 4}$ shown in Fig.~2.
Another simplification occurs by observing the symmetry of the problem under the
exchange of variables $\lambda \to 1-\lambda$: one has $\gamma_2(1-\lambda)=-\gamma_1(\lambda)$ and $\delta_2(1-\lambda)=-\delta_1(\lambda)$; here the negative sign cancels out with the reversal of the directions $\mathcal{C}_2 \to -\mathcal{C}_1$ and $\mathcal{C}_4 \to -\mathcal{C}_3$ of the contours upon reflection. Hence two pairs of the four contributions in Eq.~\eqref{eq:4cont} are equal, which yields
\begin{align}
I^{(\alpha)}_L&= \left[ \frac{a_1{-}b_1}{\pi^2} \oint_{\mathcal{C}_1} \mathrm{d} \lambda
\frac{s^{(\alpha)}(\lambda) \, 
\gamma_1(\lambda)}{(2a_1{-}1{-} \lambda)(b_1-\lambda)} +\frac{a_2{-}b_2}{\pi^2} \oint_{\mathcal{D}_1} \mathrm{d} \lambda
\frac{s^{(\alpha)}(\lambda) \, 
\delta_1(\lambda)}{(2a_2{-}1{-} \lambda)(2b_2{-}1{-}\lambda)} \right] \log L \,.
\label{eq:2cont}
\end{align}
The cuts of the functions $\gamma_1$ and $\delta_1$ are along the intervals $(2a_1-1,2b_1-1)$ and $(2a_2-1,2b_2-1)$, respectively. The jumps along these cuts can be easily calculated
\begin{align}
\gamma_1(x+\mathrm{i} 0^{\pm})=\frac{1}{2\pi i} \left[\log \frac{2a_1{-}1{-}x}{2b_1{-}1{-}x}
\mp i(\pi-0^+)\right]  =\gamma_1(x) \mp \left(\tfrac{1}{2} - 0^+\right)\,,
\;   \, x \in (2a_1{-}1,2b_1{-}1)\,, \label{cut1}
\end{align}
and similarly, 
\begin{align}
\delta_1(x+\mathrm{i} 0^{\pm})&=\delta_1(x) \mp \left(\tfrac{1}{2} - 0^+\right)\,, \;   \qquad x \in (2a_2-1,2b_2-1)\,. \label{cut2}
\end{align}
%\begin{align}
%\gamma_2(x+\mathrm{i} 0^{\pm})&=\gamma_2(x) \mp \left(\tfrac{1}{2} - 0^+\right), \;   \; x \in (1{-}b_1,1{-}a_1),\\
%\delta_1(x+\mathrm{i} 0^{\pm})&=\delta_1(x) \mp \left(\tfrac{1}{2} - 0^+\right), \;   \; x \in (a_2,b_2),\\
%\delta_2(x+\mathrm{i} 0^{\pm})&=\delta_2(x) \mp \left(\tfrac{1}{2} - 0^+\right), \;   \; x \in (1{-}b_2,1{-}a_2).
%\end{align}
%
\begin{figure}[t!]
%{\includegraphics[width=.47\textwidth]{figures/kontur1}\label{fig:cont_1}}\hfill
\center{{\includegraphics[width=.47\textwidth]{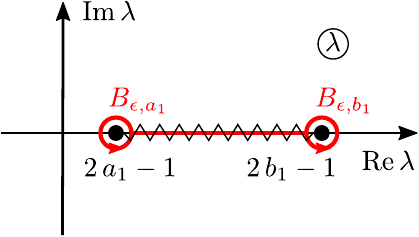}\label{fig:cont_2}}}
\caption{\label{fig:cont} The integration contour for the integrals containing $a_1$ and $b_1$ in Eq. \erf{mut_inf_cont}.}
\end{figure}
Using Eqs.~\eqref{cut1} and \eqref{cut2}, one can further decompose the contour integral along  $\mathcal{C}_1$ and $\mathcal{D}_1$ in \eqref{eq:2cont}
to integrations of the jump on the intervals $(2a_1{-}1 {+} \epsilon,2b_1{-}1 {-} \epsilon)$ and $(2a_2{-}1 {+} \epsilon, 2b_2 {-}1{-} \epsilon)$ and along circular contours around the points $2a_j-1$ and $2b_j-1$ (for 
$j=1,2$), see Fig. \ref{fig:cont}. So we obtain
\begin{align}
 I^{(\alpha)}_L &= \lim_{\epsilon \to \infty} \left[ \frac{a_1{-}b_1}{\pi^2} 
 \int_{2a_1{-}1{+}\epsilon}^{2b_1 {-}1{-} \epsilon}  \frac{ {\rm d} \lambda \,  s^{(\alpha)}(\lambda)}{(2a_1{-}1{-}\lambda)(2b_1{-}1{-}\lambda)} + \frac{a_2{-}b_2}{\pi^2}  \int_{2a_2{-}1{+}\epsilon}^{2b_2{-}1{-} \epsilon}  \frac{ {\rm d} \lambda \,  s^{(\alpha)}(\lambda)}{(2a_2{-}1{-}\lambda)(2b_2{-}1{-}\lambda)} \right. \nonumber \\
&+ \frac{a_1{-}b_1}{\pi^2}  \left(  \oint_{B_{\epsilon, a_1}} 
 \frac{\mathrm{d} \lambda \, s^{(\alpha)}(\lambda)  \gamma_1(\lambda)}{(2a_1{-}1{-} \lambda)(2b_1{-}1{-}\lambda)}  
 {+} \oint_{B_{\epsilon, b_1}}  
\frac{\mathrm{d} \lambda\, s^{(\alpha)}(\lambda)  \gamma_1(\lambda)}{(2a_1{-}1{-} \lambda)(2b_1{-}1{-}\lambda)} \right)   \nonumber \\
& + \left.  \frac{a_2{-}b_2}{\pi^2} \left(  
  \oint_{B_{\epsilon, a_2}} 
\frac{\mathrm{d} \lambda \, s^{(\alpha)}(\lambda)  \delta_1(\lambda)}{(2a_2{-}1{-} \lambda)(2b_2{-}1{-}\lambda)}  
 {+} \oint_{B_{\epsilon, b_2}}  
\frac{\mathrm{d} \lambda\, s^{(\alpha)}(\lambda)  \delta_1(\lambda)}{(2a_2{-}1{-} \lambda)(2b_2{-}1{-}\lambda)} \right) \right] \log L\,,
\label{mut_inf_cont} 
\end{align}
where $B_{\epsilon, v}$ denotes a circular contour of radius $\epsilon$ with the point $2v-1$ on the real line as the center. %One can evaluate the  four principal value integral around $v=a_1,a_2, b_1, b_2$. 
For example, for the case of $v=a_1$, after substituting $\lambda=2a_1-1+\epsilon e^{i\theta}$, one can evaluate this  principal value integral as
\begin{align}
& \lim_{\epsilon \to 0} \oint_{B_{\epsilon, a_1}} \frac{{\rm d} \lambda}{2\pi i} \, s^{(\alpha)}(\lambda) \;
\frac{\log(\lambda - 2a_1+1)-\log(\lambda- 2b_1+1)}{(\lambda -2a_1-1)(\lambda -2b_1-1)}=\nonumber \\ &  \lim_{\epsilon \to 0}  \int_{-\pi}^{\pi} \frac{{\rm d} \theta}{2\pi}\; s^{(\alpha)}(2a_1-1) \;
\frac{\log(2b_1-2a_1) -  \log (\epsilon) -i \theta}{2(b_1-a_1)}=\nonumber \\
&  \lim_{\epsilon \to 0} \, \frac{s^{(\alpha)}(2a_1-1)}{2(b_1-a_1)} \log\left(\frac{2(b_1-a_1)}{\epsilon}\right) \,.
\label{cint_1}
\end{align}
 The other principal value integrals can be calculated analogously, and we obtain the expressions
 \begin{align}
 & \lim_{\epsilon \to 0} \frac{a_1{-}b_1}{\pi^2}  \left(  \oint_{B_{\epsilon, a_1}} 
 \frac{\mathrm{d} \lambda \, s^{(\alpha)}(\lambda)  \gamma_1(\lambda)}{(2a_1{-}1{-} \lambda)(2b_1{-}1{-}\lambda)}  
 {+} \oint_{B_{\epsilon, b_1}}   
\frac{\mathrm{d} \lambda\, s^{(\alpha)}(\lambda)  \gamma_1(\lambda)}{(2a_1{-}1{-} \lambda)(2b_1{-}1{-}\lambda)} \right)=
\nonumber \\ 
&\lim_{\epsilon \to 0} \frac{s^{(\alpha)}(2a_1-1)+ s^{(\alpha)}(2b_1-1)}{2 \pi^2} \log\left(\frac{\epsilon}{2(b_1-a_1)}\right) \,,\\
& \lim_{\epsilon \to 0} \frac{a_2{-}b_2}{\pi^2}  \left(  \oint_{B_{\epsilon, a_2}} 
 \frac{\mathrm{d} \lambda \, s^{(\alpha)}(\lambda)  \delta_1(\lambda)}{(2a_2{-}1{-} \lambda)(2b_2{-}1{-}\lambda)}  
 {+} \oint_{B_{\epsilon, b_2}}   
\frac{\mathrm{d} \lambda\, s^{(\alpha)}(\lambda)  \delta_1(\lambda)}{(2a_2{-}1{-} \lambda)(2b_2{-}1{-}\lambda)} \right)=\\ 
&\lim_{\epsilon \to 0} \frac{s^{(\alpha)}(2a_2-1)+ s^{(\alpha)}(2b_2-1)}{2 \pi^2} \log\left(\frac{\epsilon}{2(b_2-a_2)}\right) \,.
\end{align}
Note that the result is divergent and the divergence is cancelled by the divergences of the line integrals.
To calculate the von Neumann mutual information $I^{(1)}_L$, we can evaluate 
the line integrals in Eq.~\eqref{mut_inf_cont} by using 
\begin{align}
&\lim_{\epsilon \to 0} \int_{2a{-}1{+}\epsilon}^{2b{-}1{-} \epsilon} \frac{ {\rm d} \lambda \,
s^{(1)}(\lambda)}{(2a{-}1{-}\lambda)(2b{-}1{-}\lambda)} =  \lim_{\epsilon \to 0} \int_{2a{-}1{+}\epsilon}^{2b{-}1{-} \epsilon}  {\rm d} \lambda \,\, \frac{-\frac{1+ \lambda}{2} \log \frac{1 + \lambda}{2} - \frac{1 - \lambda}{2} \log \frac{1- \lambda}{2}}{(2a{-}1{-}\lambda)(2b{-}1{-}\lambda)}=\nonumber \\
 &\frac{1}{2(a-b)}  \left[  \; a \, \Li{\frac{a{-}b}{a}} \right. + (1{-}a) \Li{\frac{b{-}a}{1{-}a}} 
+ b \, \Li{\frac{b{-}a}{b}}  + \left. (1{-}b) \Li{\frac{a{-}b}{1{-}b}} \,\right] \nonumber \\
 &+  \lim_{\epsilon \to 0}\frac{s^{(1)}(2a-1) + s^{(1)}(2b-1)}{2(b-a)} \log\left(\frac{\epsilon}{2(b-a)}\right)\,,
%&  \lim_{\epsilon \to 0} \int_{2a{-}1{+}\epsilon}^{2b{-}1{-} \epsilon} \frac{ {\rm d} \lambda \,
%s^{(2)}(\lambda)}{(2a{-}1{-}\lambda)(2b{-}1{-}\lambda)} =
%(s(a) + s(b)) \ln\left(\frac{\epsilon}{b-a}\right) \right]
\label{lint}
\end{align}
obtaining the final formula Eq.~\eqref{eq:sigma1_alt}.
%Adding all the terms in Eq.~\eqref{mut_inf_cont}, the final formula for the von Neumann mutual information reads

%\begin{align}
% I^{(1)}_L&=\sigma^{(1)} \log L+ const \, , \nonumber \\
%\sigma^{(1)}&=\frac{1}{2\pi^2} \sum_{i=1}^{2} \left[ a_i \, \Li{\frac{a_i{-}b_i}{a_i}} {+} (1{-}a_i) \Li{\frac{b_i{-}a_i}{1{-}a_i}}
%  \nonumber \right. \\
%& \left. \phantom{=\frac{1}{2\pi^2} \sum_{i=1}^{2}}  +  b_i \, \Li{\frac{b_i{-}a_i}{b_i}} {+}  (1{-}b_i) \Li{\frac{a_i{-}b_i}{1{-}b_i}}\right] \,. 
%\end{align}
For the R\'enyi entropy with integer $\alpha>2$ indices, we use the expression
\begin{align}
&\lim_{\epsilon \to 0}  \int_{2a{-}1{+}\epsilon}^{2b{-}1{-} \epsilon} {\rm d} \lambda \, \left[\frac{\log(\lambda-z) }{(2a{-}1{-}\lambda)(2b{-}1{-}\lambda)} + \frac{\log(\lambda-\overline{z}) }{(2a{-}1{-}\lambda)(2b{-}1{-}\lambda)} \right]= 
\nonumber \\
& \frac{1}{2(b-a)}\left[ \frac{\pi^2}{2} + {\rm Re} \left( \log^2\left( -\frac{2a{-}1{-}z}{2b{-}1{-}z}\right) + \eta\left( \frac{2a{-}1{-}z}{2b{-}1{-}z}\right)\right)\right] \nonumber\\
&+  \lim_{\epsilon \to 0}\frac{\log |2a{-}1{-}z|^2 +   \log |2b{-}1{-}z|^2}{2(b-a)} \log\left(\frac{\epsilon}{2(b-a)}\right)\,,
%(s(a) + s(b)) \ln\left(\frac{\epsilon}{b-a}\right) \right]
\end{align}
 where $z \notin \mathbb{R}$, and 
 \begin{align}  
\eta(w)=
\begin{cases}
\phantom{-}2\pi i \log (w) \; \; \text{when }  \arg(w) \in [0, \pi)\,,\\
-2\pi i \log(w) \; \; \text{when } \arg(w) \in [-\pi, 0)\,.
\end{cases}
\end{align}
Using the above line integral expression and Eq.~\eqref{cint_1}  together with the factorisations
\begin{align}
&\left(\frac{\lambda +1}{2}\right)^2 + \left(\frac{\lambda -1}{2}\right)^2= \frac{(\lambda + i) (\lambda -i)}{2}\,,\\
&\left(\frac{\lambda +1}{2}\right)^3 + \left(\frac{\lambda -1}{2}\right)^3=\frac{3(\lambda + i/\sqrt{3}) (\lambda -i/\sqrt{3})}{8}\,,\\
&\left(\frac{\lambda +1}{2}\right)^4 + \left(\frac{\lambda -1}{2}\right)^4= \frac{\left(\lambda {+} i \tan \frac{\pi}{8}\right)\left(\lambda {+} i \tan \frac{3\pi}{8}\right)\left(\lambda {+} i \tan \frac{5\pi}{8}\right)\left(\lambda {+} i \tan \frac{7\pi}{8}\right)}{8}\,,\\
&\left(\frac{\lambda +1}{2}\right)^{2^m} + \left(\frac{\lambda -1}{2}\right)^{2^m}= \frac{1}{2^{2^m-1}}\prod_{k=1}^{2^{m-1}}\left(\lambda + i \tan \frac{(2k-1)\pi i}{2^{m+1}}\right)\,,
\end{align}
we can evaluate the integral \eqref{mut_inf_cont} for $\alpha=2,3,4,2^m$ and obtain the results stated in Eq. \eqref{sigmas}.

\section*{References}

\bibliographystyle{iopart-num}

%\bibliography{Renyi_refs}
\bibliography{IsingNegRenyi_resub}

\end{document}